\documentclass[fleqn,usenatbib]{mnras}

\usepackage{newtxtext}

\usepackage[T1]{fontenc}

\usepackage[usenames,dvipsnames]{xcolor}

\DeclareRobustCommand{\VAN}[3]{#2}
\let\VANthebibliography\thebibliography
\def\thebibliography{\DeclareRobustCommand{\VAN}[3]{##3}\VANthebibliography}

\usepackage[normalem]{ulem}
\usepackage{subfig}
\usepackage{graphicx}	
\usepackage{amsmath}	
\usepackage{amssymb}	
\usepackage{url}
\usepackage{balance}

\usepackage[most]{tcolorbox}
\usepackage{tikz,lipsum,lmodern}

\usepackage{mathtools}
\usepackage{siunitx}

\usepackage{fourier} 
\usepackage{array}
\usepackage{makecell}

\usepackage{orcidlink}

\usepackage{hyperref}

\newcommand\blfootnote[1]{%
  \begingroup
  \renewcommand\thefootnote{}\footnote{#1}%
  \addtocounter{footnote}{-1}%
  \endgroup
}

\hypersetup{
    colorlinks = true,
    citecolor = {MidnightBlue},
    linkcolor = {BrickRed},
    urlcolor = {BrickRed}
}

\DeclarePairedDelimiterXPP\BigOSI[2]%
  {\mathcal{O}}{(}{)}{}%
  {\SI{#1}{#2}}
\DeclareMathAlphabet{\mathcal}{OMS}{cmsy}{m}{n}
\SetMathAlphabet{\mathcal}{bold}{OMS}{cmsy}{b}{n}

\DeclareRobustCommand{\orderof}{\ensuremath{\mathcal{O}}}
\title[Bayesian estimation of systematics in 21cm data]{Bayesian estimation of cross-coupling and reflection systematics \\
in 21cm array visibility data}

\author[G. Murphy et al.]{Geoff G. Murphy\,\orcidlink{0000-0002-8186-3064}$^{1}$,
	Philip  Bull\,\orcidlink{0000-0001-5668-3101}$^{2,1}$,
	Mario G. Santos\,\orcidlink{0000-0003-3892-3073}$^{1,3}$,
	Zara  Abdurashidova$^{4}$,
	Tyrone  Adams$^{3}$,
\newauthor
	James E. Aguirre\,\orcidlink{0000-0002-4810-666X}$^{5}$,
	Paul  Alexander$^{6}$,
	Zaki S. Ali$^{4}$,
	Rushelle  Baartman$^{3}$,
	Yanga  Balfour$^{3}$,
\newauthor
	Adam P. Beardsley\,\orcidlink{0000-0001-9428-8233}$^{7,8}$,
	Gianni  Bernardi\,\orcidlink{0000-0002-0916-7443}$^{9,10,3}$,
	Tashalee S. Billings$^{5}$,
	Judd D. Bowman\,\orcidlink{0000-0002-8475-2036}$^{7}$,
\newauthor
	Richard F. Bradley$^{11}$,
	Jacob  Burba\,\orcidlink{0000-0002-8465-9341}$^{2}$,
	Christopher Cain$^{7}$,
	Steven  Carey$^{6}$,
	Chris L. Carilli\,\orcidlink{0000-0001-6647-3861}$^{12}$,
\newauthor
	Carina  Cheng$^{4}$,
	David R. DeBoer\,\orcidlink{0000-0003-3197-2294}$^{13}$,
	Eloy  de~Lera~Acedo$^{6}$,
	Matt  Dexter$^{13}$,
	Joshua S. Dillon\,\orcidlink{0000-0003-3336-9958}$^{4}$,
\newauthor
	Nico  Eksteen$^{3}$,
	John  Ely$^{6}$,
	Aaron  Ewall-Wice\,\orcidlink{0000-0002-0086-7363}$^{4,14}$,
	Nicolas  Fagnoni$^{6}$,
	Randall  Fritz$^{3}$,
\newauthor
	Steven R. Furlanetto\,\orcidlink{0000-0002-0658-1243}$^{15}$,
	Kingsley  Gale-Sides$^{6}$,
	Brian  Glendenning$^{16}$,
	Deepthi  Gorthi\,\orcidlink{0000-0002-0829-167X}$^{4}$,
\newauthor
	Bradley  Greig\,\orcidlink{0000-0002-4085-2094}$^{17}$,
	Jasper  Grobbelaar$^{3}$,
	Ziyaad  Halday$^{3}$,
	Bryna J. Hazelton\,\orcidlink{0000-0001-7532-645X}$^{18,19}$,
\newauthor
	Jacqueline N. Hewitt\,\orcidlink{0000-0002-4117-570X}$^{20,21}$,
	Jack  Hickish$^{13}$,
	Daniel C. Jacobs\,\orcidlink{0000-0002-0917-2269}$^{7}$,
	Austin  Julius$^{3}$,
	MacCalvin  Kariseb$^{3}$,
\newauthor
	Nicholas S. Kern\,\orcidlink{0000-0002-8211-1892}$^{4,21}$,
	Joshua  Kerrigan\,\orcidlink{0000-0002-1876-272X}$^{22}$,
	Piyanat  Kittiwisit\,\orcidlink{0000-0003-0953-313X}$^{1}$,
	Saul A. Kohn\,\orcidlink{0000-0001-6744-5328}$^{5}$,
	Matthew  Kolopanis\,\orcidlink{0000-0002-2950-2974}$^{7}$,
\newauthor
	Adam  Lanman$^{22}$,
	Paul  La~Plante\,\orcidlink{0000-0002-4693-0102}$^{4,5}$,
	Adrian  Liu\,\orcidlink{0000-0001-6876-0928}$^{23}$,
	Anita  Loots$^{3}$,
	David Harold~Edward MacMahon$^{13}$,
\newauthor
	Lourence  Malan$^{3}$,
	Cresshim  Malgas$^{3}$,
	Keith  Malgas$^{3}$,
	Bradley  Marero$^{3}$,
	Zachary E. Martinot$^{5}$,
\newauthor
	Andrei  Mesinger\,\orcidlink{0000-0003-3374-1772}$^{24}$,
	Mathakane  Molewa$^{3}$,
	Miguel F. Morales\,\orcidlink{0000-0001-7694-4030}$^{18}$,
	Tshegofalang  Mosiane$^{3}$,
\newauthor
	Steven G. Murray\,\orcidlink{0000-0003-3059-3823}$^{24,7}$,
	Abraham R. Neben\,\orcidlink{0000-0001-7776-7240}$^{21}$,
	Bojan  Nikolic$^{6}$,
	Hans  Nuwegeld$^{3}$,
	Aaron R. Parsons\,\orcidlink{0000-0002-5400-8097}$^{4}$,
\newauthor
	Nipanjana  Patra\,\orcidlink{0000-0002-9457-1941}$^{4}$,
	Samantha  Pieterse$^{3}$,
	Nima  Razavi-Ghods$^{6}$,
	James  Robnett$^{12}$,
\newauthor
	Kathryn  Rosie$^{3}$,
	Peter  Sims\,\orcidlink{0000-0002-2871-0413}$^{23}$,
	Jackson Sipple$^{4}$,
	Craig  Smith$^{3}$,
	Hilton  Swarts$^{3}$,
\newauthor
	Nithyanandan  Thyagarajan\,\orcidlink{0000-0003-1602-7868}$^{25}$,
	Pieter  van~Wyngaarden$^{3}$,
	Peter K.~G. Williams\,\orcidlink{0000-0003-3734-3587}$^{26,27}$,
	Haoxuan  Zheng$^{21}$
\\
$^{1}$ Department of Physics and Astronomy,  University of the Western Cape, Cape Town, 7535, South Africa\\
$^{2}$ Jodrell Bank Centre for Astrophysics, University of Manchester, Manchester, M13 9PL, United Kingdom\\
$^{3}$ South African Radio Astronomy Observatory, Black River Park, 2 Fir Street, Observatory, Cape Town, 7925, South Africa\\
$^{4}$ Department of Astronomy, University of California, Berkeley, CA\\
$^{5}$ Department of Physics and Astronomy, University of Pennsylvania, Philadelphia, PA\\
$^{6}$ Cavendish Astrophysics, University of Cambridge, Cambridge, UK\\
$^{7}$ School of Earth and Space Exploration, Arizona State University, Tempe, AZ\\
$^{8}$ Department of Physics, Winona State University, Winona, MN\\
$^{9}$ INAF-Istituto di Radioastronomia, via Gobetti 101, 40129 Bologna, Italy\\
$^{10}$ Department of Physics and Electronics, Rhodes University, PO Box 94, Grahamstown, 6140, South Africa\\
$^{11}$ National Radio Astronomy Observatory, Charlottesville, VA\\
$^{12}$ National Radio Astronomy Observatory, Socorro, NM 87801, USA\\
$^{13}$ Radio Astronomy Lab, University of California, Berkeley, CA\\
$^{14}$ Department of Physics, University of California, Berkeley, CA\\
$^{15}$ Department of Physics and Astronomy, University of California, Los Angeles, CA\\
$^{16}$ National Radio Astronomy Observatory, Socorro, NM\\
$^{17}$ School of Physics, University of Melbourne, Parkville, VIC 3010, Australia\\
$^{18}$ Department of Physics, University of Washington, Seattle, WA\\
$^{19}$ eScience Institute, University of Washington, Seattle, WA\\
$^{20}$ MIT Kavli Institute, Massachusetts Institute of Technology, Cambridge, MA\\
$^{21}$ Department of Physics, Massachusetts Institute of Technology, Cambridge, MA\\
$^{22}$ Department of Physics, Brown University, Providence, RI\\
$^{23}$ Department of Physics and Trottier Space Institute, McGill University, 3600 University Street, Montreal, QC H3A 2T8, Canada\\
$^{24}$ Scuola Normale Superiore, 56126 Pisa, PI, Italy\\
$^{25}$ Commonwealth Scientific and Industrial Research Organisation (CSIRO), Space \& Astronomy, P. O. Box 1130, Bentley, WA 6102, Australia\\
$^{26}$ Center for Astrophysics, Harvard \& Smithsonian, Cambridge, MA\\
$^{27}$ American Astronomical Society, Washington, DC
}

\date{Accepted XXX. Received YYY; in original form ZZZ}

\pubyear{2023}

\begin{document}
\label{firstpage}
\pagerange{\pageref{firstpage}--\pageref{lastpage}}
\maketitle

\clearpage
\begin{abstract}
Observations with radio arrays that target the 21-cm signal originating from the early Universe suffer from a variety of systematic effects. An important class of these are reflections and spurious couplings between antennas. We apply a Hamiltonian Monte Carlo sampler to the modelling and mitigation of these systematics in simulated Hydrogen Epoch of Reionisation Array (HERA) data. This method allows us to form statistical uncertainty estimates for both our models and the recovered visibilities, which is an important ingredient in establishing robust upper limits on the Epoch of Reionisation (EoR) power spectrum. In cases where the noise is large compared to the EoR signal, this approach can constrain the systematics well enough to mitigate them down to the noise level for both systematics studied. Incoherently averaging the recovered power spectra can further reduce the noise and improve recovery. Where the noise level is lower than the EoR, our modelling can mitigate the majority of the reflections and coupling with there being only a minor level of residual systematics. Our approach performs similarly to existing filtering/fitting techniques used in the HERA pipeline, but with the added benefit of rigorously propagating uncertainties. In all cases it does not significantly attenuate the underlying signal.
\end{abstract}

\begin{keywords}
methods: statistical, data analysis -- techniques: interferometric -- cosmology: dark ages, reionization, first stars
\end{keywords}



\section{Introduction}\label{section:Introduction}
\blfootnote{Corresponding authors: Philip Bull (phil.bull@manchester.ac.uk), Geoff Murphy (4178310@myuwc.ac.za)}The Hydrogen Epoch of Reionisation Array (HERA) is an interferometer located in the Karoo desert in South Africa. It was purpose-built to detect the statistical fluctuations in the brightness temperature of the redshifted 21-cm radio emission from the Epoch of Reionisation \citep[EoR,][]{DeBoer_2017}. This signal is emitted by neutral hydrogen during a period when the first stars and galaxies were formed and began to ionise the intergalactic medium (IGM). Depending on the model, the Epoch of Reionisation (EoR) is expected to occur over a redshift of $6 \lesssim z \lesssim 10$ \citep{Furlanetto_2006,Liu+20}. Measuring the spatial and temporal evolution of this signal promises to greatly improve our understanding of the first gravitationally-bound objects formed in the Universe, as well as their surrounding environment \citep{Kern_2017, Ewall-Wice_2016a}. A number of other experiments have also attempted to measure this 21-cm signal, including the Donald C. Backer Precision Array for Probing the Epoch of Reionization \citep[PAPER,][]{Parsons_2010,Kolopanis_2019}, the Giant Meter Wave Radio Telescope \citep[GMRT,][]{Swarup_1991,Paciga_2013}, the Low Frequency Array \citep[LOFAR,][]{vanHaarlem+13,Mertens+20}, the Long Wavelength Array \citep[LWA, ][]{Eastwood+19,Garsden+21}, and the Murchison Widefield Array \citep[MWA,][]{tingay_2013,Rahimi_2021}.

Recently, the upper limits on the EoR power spectrum at $z=7.9$ and $z=10.4$ have been lowered to their most sensitive yet using HERA Phase I data \citep{Abdurashidova_2023}, obtained by observing over 94 nights with 35 to 41 antennas. These limits are an improvement of over a factor of two compared to the results in \cite{Abdurashidova_2022b}. At $z=7.9$, the upper limit on the power spectrum has been placed at $\Delta^2(k=0.34\,h\rm{Mpc}^{-1}) \le 457\,\rm{mK}^2$, and $\Delta^2(k=0.36\,h\rm{Mpc}^{-1}) \le 3,496 \,\rm{mK}^2$ at $z=10.4$. Using analysis methods from \cite{Abdurashidova_2022a}, this allowed for a number of inferences on early-Universe properties and models. For example, the results suggest that the IGM had to be at least slightly heated prior to reionisation. This is contrary to what "cold-reionisation" models propose, which is a much cooler IGM before the majority of reionisation occurs. It is expected that high-mass X-ray binaries (HMXBs) are the primary source of this heating \citep{Fragos_2013}. HERA's newest results agree best with low-metallicity HMXB models, as opposed to the high-metallicity counterparts.

The data modelled here corresponds to HERA Phase I, where observations were carried out with 35 to 41 antennas, and which repurposed PAPER's dipoles, correlator, and signal chains \citep{Abdurashidova_2023}. A fully constructed Phase II array will consist of 350 dishes, replaces the correlator and signal chain, and has Vivaldi feeds which increase the observing bandwidth \citep{Fagnoni_2021}. This should result in more sensitive upper limits, and improved constraints on early-Universe models, with the ultimate goal being a direct detection of the EoR 21-cm signal. Forecasts suggest that a fully constructed HERA should be sensitive enough to make a $\gtrsim30\sigma$ detection of the EoR power spectrum when using even fairly basic analysis techniques. More sophisticated analysis (foreground removal, for example) could see an even greater increase in sensitivity \citep{Pober_2014}.

However, given the faintness of the 21-cm signal, and by extension the sensitivity required for measurement, a great deal of work has been put into understanding the effects of performance-limiting effects. Examples include radio frequency interference \citep[RFI,][]{Wilensky_2020}, errors in redundant calibration \citep{Byrne_2019,Orosz_2019}, the effects of data flagging and in-painting \citep{Offringa_2019,Pagano_2023}, and absolute calibration errors \citep{Kern_2020b}. These, as well as other effects can result in a loss of the 21-cm signal if not fully accounted for. For the analysis pipeline of \cite{Abdurashidova_2023}, redundant-baseline averaging results in a $1.9\%$ and $2.4\%$ signal loss in frequency bands 1 and 2, respectively, due to slight non-redundancies and the corresponding decoherences. Redundant time averaging results in an additional $1.2\%$ and $1.5\%$ signal loss due to the slight differences in sky signal between integrations.

Most relevant to this work, however, is the effect of internal instrument coupling. Fig. \ref{fig:hera_illust} shows a schematic of the systematics in HERA Phase I which are relevant to this work. As it is defined in \cite{Kern+19}, these systematics are signal chain reflections, and antenna cross-coupling. Reflections result from a portion of the measured signal being reflected at either end of the coaxial cables connecting the various components in the array, for example the connections between antennas and correlators and/or digitisers. Cable wear can also result in multiple reflections as a signal travels along the cable, which is referred to as subreflections. In order to move the reflection systematic towards higher delay modes (away from those of scientific interest), HERA Phase II lengthens the connection between the Front End Module (FEM) and Post Amplifier Module (PAM) with a 500\,m fiber cable \citep{Berkhout_2024}.

In 21-cm experiments, multiple systematics exist which act to couple antennas to one another, for example stray capacitance between adjacent wires, and over-the-air antenna-to-antenna reflections \citep[mutual coupling,][]{Josaitis_2022}. In this work, however, we focus only on the form of coupling seen in HERA Phase I, referred to as cross-coupling. This systematic was hypothesised to be a result of broadcasting signals originating from a leaking connection point. This unintentionally broadcast voltages measured by the array, where it was measured again by the antennas, resulting in copies of the data being added to the overall data stream. However, the equipment responsible for this particular systematic is no longer in use, and so this systematic is not present in recent data \citep{Abdurashidova_2023}. Nevertheless, demonstrating that signal chain systematics such as this can be modelled with a Bayesian approach is useful for future Phase II observations.

While foregrounds (mainly Milky Way emission) are up to $10^5$ times stronger than the EoR signal in observations, they are ideally confined to the so-called foreground wedge, leaving a foreground-free window \citep{Vedantham_2012,Parsons_2012a,Parsons_2012b,Liu_2014a,Liu_2014b}. However, these systematics spread the foreground signal into the window, overwhelming the EoR signal. The aim is to model these systematics so that they can be removed from the observational data, allowing for a recovery of the underlying EoR signal.

For the systematics of interest in this work, the current removal strategies have been shown to be effective. \cite{Kern+19} model and simulate reflections and cross-coupling in 21-cm observations and demonstrate associated removal strategies. They find that, in realistic cases where there are many nearly overlapping systematics in a single visibility, it is possible to mitigate these features while avoiding any significant attenuation of the underlying 21-cm signal. In \cite{Kern+20}, these methods are applied to early HERA Phase I data \citep{DeBoer_2017}, and are capable of recovering signal down to the noise floor. 
These methods for reflection and cross-coupling removal are still in use for the latest data \citep{Abdurashidova_2023}. 

This paper aims to extend on the current systematic mitigation techniques by modelling them in a Bayesian framework, which naturally provides a measure of the statistical uncertainty. The intention is that these uncertainties will be propagated into results further down the analysis chain, providing a level of confidence that EoR signal has not been inadvertently attenuated in observational data and rigorously reflecting the additional uncertainties on the recovered signal arising from the systematics removal procedure.

The paper is organised as follows: Section\,\ref{section:Systematics} describes the systematics of interest in this work, and how they are currently removed from observational data. Section\,\ref{section:Data} details the data used, and the characteristics of the systematics added, while Section\,\ref{section:Modelling} discusses our foreground model, Monte Carlo sampler setup, the systematics subtraction method, and the signal loss metric. Section\,\ref{section:Results} presents the results, and Section\,\ref{section:Discussion} summarises the work.

\section{Systematic effects}\label{section:Systematics}
Three systematics present in the HERA Phase I instrument are modelled in this work. These are cable reflections, cable subreflections, and cross-coupling arising from a leaking connection point. These particular systematics have been modelled in previous works, namely \citet{Kern+19} and \citet{Aguirre+22}, and so have equations which describe their expected form in observational data. As detailed in \cite{Kern+19} and \cite{Kern+20}, these systematics are also expected to be essentially time-stable. Cross-coupling systematics vary on the order of around 1 hr, and so can be treated as having constant amplitudes in HERA analysis which time-averages visibilities to a cadence of 21.4\,s prior to systematics removal \citep{Abdurashidova_2023}. The cross-coupling delays (dependent on the antenna positions) and phases are not expected to be time-varying. 

\begin{figure*}
    \centering
    \includegraphics[width=1.8\columnwidth]{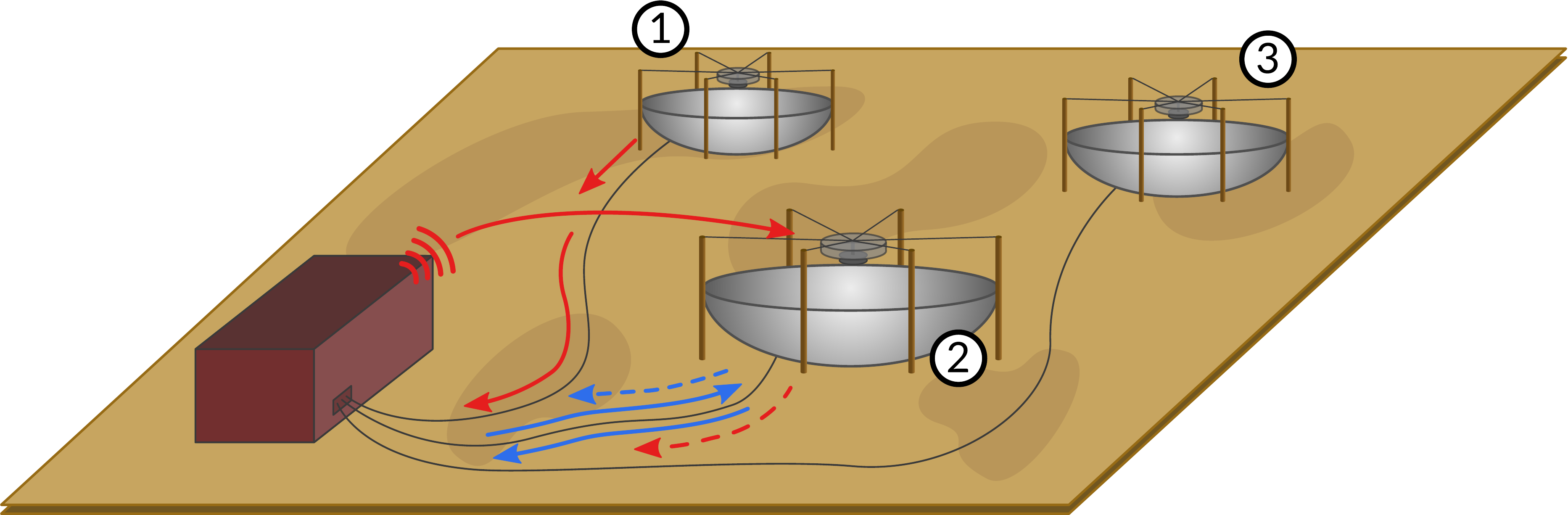}
    \caption{An illustration of the sources of the cable reflections and cross-coupling systematics in HERA. Antennas are connected via coaxial cables to housings next to the array. The solid blue line travelling towards the antenna denotes the reflecting signal in the cable, with the dashed blue line indicating the copy which is added to the overall data stream. The red solid line denotes the source of the cross-coupling. The sky signal measured by antenna 1 travels down the signal chain to the equipment in the housing, thereafter to a faulty connection point which was the likely source of the broadcasting signal. Antenna 2 picks up this broadcasted signal, resulting in a copy of antenna 1's autocorrelation being added to antenna 2's visibility.}
    \label{fig:hera_illust}
\end{figure*}

While the conditions which give rise to reflections are unchanging, for example the length of the cables, these systematics are expected to vary slightly as a function of the foregrounds. However, for the time-averaging cadences used in HERA analyses, reflections are also treated as time-stable. This, and the fact that these systematics can all be described by only three free parameters per individual feature, means that they lend themselves well to forward-modelling. Fig. \ref{fig:hera_illust} provides a diagrammatic description of the sources of these cable reflections and cross-coupling.

All the considered systematics insert copies of observed signals into the data, but at regions in Fourier space where they are not normally found. The foregrounds dominate these copies, and are the primary components which overwhelm the 21-cm signal. The EoR signal itself, as well as internal instrument coupling systematics already present in the signal chain, are copied as well, but these have amplitudes weaker than the 21-cm signal, and so are not a concern as they are second order effects. While the systematics amplitudes can be up to six orders of magnitude smaller than the foreground peak (for the power spectra in this paper), even this is enough to overwhelm the 21-cm signal in regions of Fourier space which are of most interest. This contamination of ideally foreground-free regions violates HERA's foreground avoidance strategy \citep{Kerrigan_2018,Morales_2018}, which attempts to localise all foregrounds to the foreground wedge. Furthermore, poor attempts to remove these systematics and recover the true sky signal can result in an over-subtraction, attenuating the already weak 21-cm power spectrum.

A number of radio and 21-cm experiments have been impacted by similar systematics, necessitating a thorough understanding of their sources and mitigation. In particular, early MWA data suffered from cable reflection systematics which limited its sensitivity to the 21-cm signal from the Epoch of X-ray heating at a redshift of $12 \lesssim z \lesssim 18$ \citep{Ewall-Wice_2016}, and the Epoch of Reionisation \citep[$z=7.1$,][]{Beardsley_2016}. By including a reflection term into the gain solutions, it was possible to mitigate these reflections \citep{Jacobs_2016}. 

The low-frequency components of the Square Kilometre Array (SKA1-Low) were designed with mutual coupling in mind. Mutual coupling refers to an over-the-air, antenna-to-antenna reflection effect, and it is expected that HERA will suffer from this systematic in future (see Section \ref{subsection:cross_coupling}). Despite the fact that the level of mutual coupling is expected to be higher for the SKA than for HERA, its effects in the SKA are reduced when the data is averaged by placing the antennas in pseudo-random positions \citep{de_Lera_Acedo_2017}. Nevertheless, it is still expected to be a concern in future SKA1-Low observations, for example at the calibration step \citep{Borg_2020}.

The visibility measured by a baseline (\textbf{b}) formed by two antennas, and for a given frequency ($\nu$), is 

\begin{equation}
    V(\rm{\textbf{b}},\nu) = \int d^2\theta\, T(\pmb{\theta},\nu)\, A_p(\pmb{\theta},\nu)e^{-i2\pi\textbf{b}\cdot \pmb{\theta}/\lambda} .
\end{equation}

$\rm{T}(\pmb{\theta})$ is the sky brightness temperature, $\rm{A}_{\rm{p}}(\pmb{\theta})$ the primary beam, $\pmb{\theta}$ the coordinates on the sky in the flat-sky approximation, where $\pmb{\theta}\equiv(\theta_x,\theta_y)$, and $\lambda$ the wavelength of the radiation measured. Most relevant is the Fourier transform (or delay transform) of this visibility: 

\begin{equation}
    \Tilde{V}_b(\tau) \equiv \int d\nu V(\nu)\phi(\nu)e^{-i2\pi\nu\tau},
\end{equation}

where the delay ($\tau$) is the Fourier dual of frequency and has units of time, and $\phi(\nu)$ is a tapering function \citep{Liu+20}. 

The delay quantity is of interest in this work as relevant signal chain systematics create copies of incoming radiation which arrive after the primary radiation signal. This time delay of the systematics is dependent on the geometry of the array, for example the lengths of the connecting cables, or the distance between antennas.

\subsection{Cable Reflections}\label{subsection:CR}

Fig. \ref{fig:hera_illust} demonstrates the source of cable reflections with blue arrows. Following signal collection by a HERA antenna, signal should ideally travel away from said antenna and towards the amplifiers, digitizers, correlators, etc. Instead, a portion of this signal reflects at the end of the connecting cable, travels back towards the antenna, reflects again (blue dashed arrow), and is finally added to the overall data stream. This can occur for each cable connecting an antenna. 

\begin{figure*}
    \centering
    \includegraphics[width=2\columnwidth]{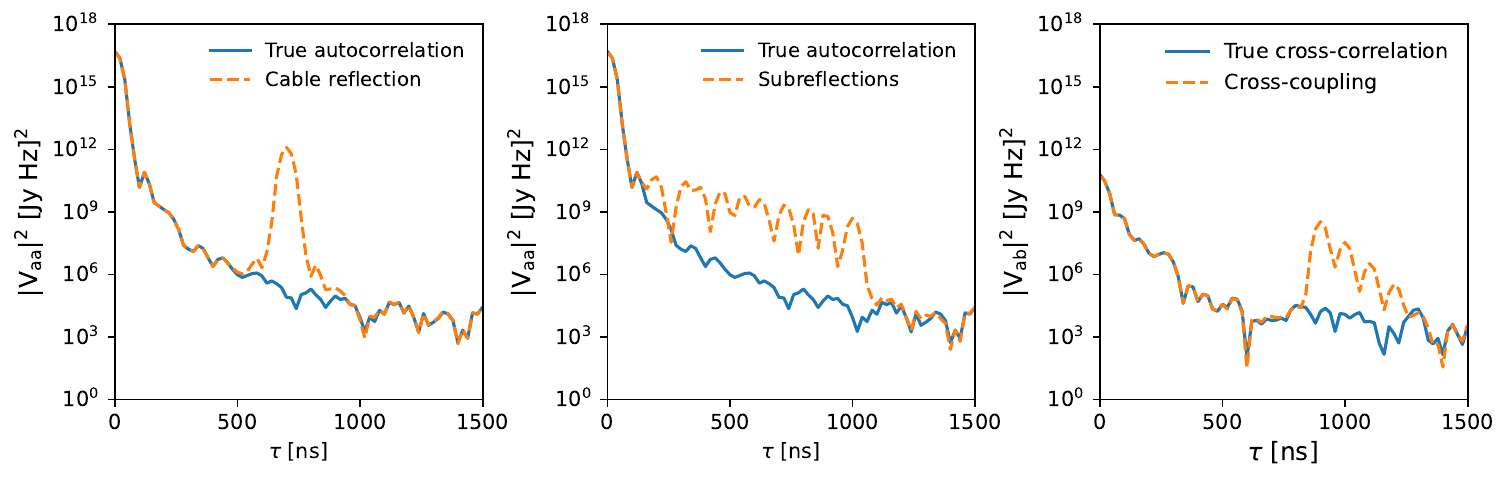}
    \caption{Left: An example of a cable reflection in an autocorrelation power spectrum. The true spectrum is the blue solid line, and the corrupted spectrum is dashed-orange. The cable reflection was simulated with a delay of $700\:\rm{ns}$ and a relative amplitude of $5\times10^{-3}$. The delay ($\tau$) is the Fourier dual of frequency. Middle: A demonstration of ten subreflections inserted (non-uniformly) between $200-1000\:\rm{ns}$, with amplitudes ranging from $10^{-3}-10^{-4}$, for an autocorrelation power spectrum. Right: An example of five cross-coupling peaks added to a cross-correlation power spectrum, inserted at $900-1300\:\rm{ns}$, with relative amplitudes ranging from $10^{-4}-10^{-6}$. }
    \label{fig:Sys_demo}
\end{figure*}

This reflected signal's amplitude is lower than the incident signal, has a possible phase shift, and is delayed in time. This delay is equal to twice the cable length divided by the speed of light in the cable. Fig. \ref{fig:Sys_demo} (left) demonstrates the effect of a single cable reflection with arbitrary amplitude and delay in a simulated autocorrelation power spectrum. The true data contains only foregrounds and the EoR, but the reflection systematic adds a copy of this power spectrum at higher delays. While the power of this copy is significantly lower than the true data, the relative strength of the foregrounds result in this nevertheless overwhelming the EoR signal.

As presented in \citep{Kern+19}, reflections can be described with a coupling coefficient. In this case, reflections couple signals with themselves:
\begin{equation}\label{eq:epsilon}
    \epsilon_{11}(\nu) = \rm{A}_{11} e^{2\pi i \tau_{11}\nu+i\phi_{11}}.
\end{equation}
Here, $\rm{A}$ refers to the relative amplitude of the reflection in comparison to the foreground peak of the corrupted visibility, $\tau$ is its time delay, and $\phi$ is its phase offset, all for antenna~1. A reflection-corrupted cross-correlation visibility can be described as 
\begin{equation}\label{CR_cross_coupling}
    V'_{12} = v_1v_2^* + \epsilon_{11}v_1v_2^* + v_1\epsilon_{22}^*v_{2}^* + \epsilon_{11}v_1\epsilon_{22}^*v_2^* ,
\end{equation}
where the first term is the fiducial cross-correlation visibility formed by each antenna's respective voltage spectrum. The second and third terms are copies of this visibility, arising from antenna 1 (at positive delays) and antenna 2 (at negative delays), respectively. The final term is a second-order effect. Depending on the strength of the coupling coefficient, this may produce an additional non-negligible systematic feature at higher delays, or it could result in a systematic which is weaker than the 21-cm signal.

For the systematics simulated in this work, first-order reflections begin at $200$\:ns. With relative amplitudes of $\sim10^{-3}$ (in visibility-space), second-order reflections would begin at $400$\:ns, and would be slightly stronger than the foregrounds and signal. Similar can be said of the power of the second-order coupling systematics. It would be fairly straightforward to include second-order reflections, as the delays are simply dependent on both antenna one and antenna two's cable lengths, and the relative amplitudes would be the product of $A_{11}$ and $A_{22}$. However, noise and imperfect knowledge of the instrument would complicate matters, so distinct parameters for first and second-order effects would possibly be required. It is certainly feasible, though, to use the model estimates for the first-order systematics to estimate the second-order parameters, rather than sampling for the latter. At the very least the relation between the different orders can be used to inform the priors. We opt to neglect the second-order terms, however, in order to avoid additional parameters in an already high-dimensional model.

Reflections produce a similar effect in the autocorrelations, but instead only add visibility copies of the singular antenna in question,
\begin{equation}
    V_{11}' = v_1v_1^* + \epsilon_{11}v_1v_1^* + v_1\epsilon_{11}^*v_1^* + |\epsilon_{11}|^2v_1v_1^*.
\end{equation}
Here, copies of the same autocorrelation visibility are added at both positive and negative delays. 

Following the description of \cite{Aguirre+22}, the total effect of $\rm{M}$ cable reflections in a particular signal chain are modelled as the product of the respective gain terms, 

\begin{equation}\label{eq:CableReflectionGain}
    \tilde{g}_a = \prod\limits_{j}^{\rm{M}}(1 + {\rm{A}}_{a,j}e^{i2\pi\nu\tau_{a,j} + i\phi_{a,j}}) .
\end{equation}

Here, antenna $a$ is corrupted by $\rm{M}$ reflections. Gain terms account for unintended variations in phase and amplitude introduced by the signal chain between the antenna and correlator \citep{Kern+19}. In this case, it can be used to correct for the excess amplitude and phase shifts produced by cable reflections.

For a given uncorrupted visibility, the reflection gains are applied with

\begin{equation}\label{eq:CableReflectionApply}
    V_{apbq,t}^{\rm{refl}} = \tilde{g}_a\tilde{g}_b^*V_{apbq,t}^{} \:,
\end{equation}
where $V_{apbq,t}^{}$ is the true visibility \citep{Aguirre+22}. This refers to the visibility formed by antennas $a$ and $b$, with feeds $p$ and $q$, respectively, across an integration time of $t$.

Since reflections couple signals with themselves, and autocorrelations have no imaginary component, in delay-space this systematic is symmetric around $\tau=0$. This is not necessarily the case for the cross-correlations, as the two first-order terms in Eq. \ref{CR_cross_coupling} differ by their coupling coefficients, $\epsilon$. For each coefficient, the specific amplitudes, delays, and phases can, in turn, differ for each antenna, resulting in distinct features on either side of $\tau=0$.

\subsection{Cable Subreflections}
Imperfections, wear, and damage to connecting cables can result in additional copies of a signal being added to the data stream. In this case, copies are made of a signal as it travels along the cable, rather than being a result of reflections at either end \citep{Kern+20}. These subreflections are once again modelled as reflection terms (i.e. using Eqs. \ref{eq:CableReflectionGain} and \ref{eq:CableReflectionApply}), but are more numerous, have lower amplitudes, and are spread across a range of delays. These, too, are a per-antenna effect. Fig. \ref{fig:Sys_demo} (middle) provides an example of how these subreflections affect an autocorrelation power spectrum.

\subsection{Cross-coupling}\label{subsection:cross_coupling}

Cross-coupling has multiple sources, but for the purposes here, we focus on over-the-air effects. This includes the systematic present in HERA Phase I, which was a faulty connection point, as well as mutual coupling wherein signal from the sky is reflected by one antenna and is in turn measured by another. This, again, adds a copies of visibilities to the data stream. In the simplistic, two-antenna regime, the interferometric visibility has autocorrelations from both antennas added, as well as a copy of the cross-correlation (although this is a second-order term). As described in \cite{Kern+19}, the corrupted cross-correlation of antennas 1 and 2 (denoted by $V'$) is 
\begin{equation}
    V'_{12} = v_1 v_2^* + v_1\epsilon^*_{12}v_1^* + \epsilon_{21}v_2v_2^* + \epsilon_{21}v_2\epsilon_{12}^*v_1^*.
    \label{eq:Cross_coupling}
\end{equation}
The first term is simply the uncorrupted cross-correlation. The second and third terms are the autocorrelations for antennas 1 and 2, respectively, both multiplied by the associated coupling coefficient. For $\epsilon_{12}$, this describes the amplitude and delay of antenna 2's voltage when coupled with antenna 1's voltage. The fourth term is the second-order effect which can be neglected. 

\cite{Kern+19} model $\epsilon$ as a type of reflection systematic, for simplicity. As with reflections, the second term of Eq. \ref{eq:Cross_coupling} states that antenna 1's autocorrelation is copied at positive delays in $V_{12}'$, while the third term is that of antenna 2, copied at negative delays. 

Autocorrelation data suffers from cross-coupling systematics as well. From \cite{Kern+19}, the corrupted autocorrelation visibility of antenna 1 is
\begin{equation}
    V'_{11} = v_1 v_1^* + v_1\epsilon^*_{21}v_2^* + \epsilon_{21}v_2v_1^* + |\epsilon_{21}|^2 v_2 v_2^* .
    \label{eq:Cross_coupling_auto}
\end{equation}
Here, the leading order terms contain a copy of the cross-correlation between antennas 1 and 2, while the copied autocorrelation of antenna 2 is of second order. However, cross-correlations are typically lower in power than the corresponding autocorrelations. With the additional amplitude suppression from $\epsilon_{21}$ and $\epsilon_{12}$, this results in cross-coupling systematics being significantly weaker in autocorrelations. 

Hence, to first order, cross-coupling adds copies of the autocorrelations to the cross-correlation visibilities, and copies of the cross-correlations to the autocorrelations.

Fig. \ref{fig:Sys_demo} (right) provides a demonstration for this final systematic in a cross-correlation power spectrum. The appearance would look similar in an autocorrelation, although the dynamic range between the foregrounds and cross-coupling peaks would be larger.

The second possible source of cross-coupling is capacitive crosstalk, wherein signal chains in close proximity to one another interact electromagnetically, inducing voltages. This systematic can be mitigated at the hardware level \citep{Chaudhari_2017}, and so this source of coupling is not a concern in the current analysis, nor in \cite{Kern+19}, \cite{Kern+20}, or \cite{Aguirre+22}.

In HERA Phase I observational data, mutual coupling (the antenna-to-antenna reflections) was not of major concern, either. As was detailed previously, coupling systematics in Phase I were attributed to a leaking connection point, which broadcast the voltage signals of antennas in the array, and which were in turn measured again by the antennas \citep{Abdurashidova_2023}. This has since been fixed, so this particular source is no longer an issue. However, with an increase in the number of antennas as HERA nears completion, as well as the introduction of Vivaldi feeds for these antennas, mutual coupling considerations will become more important. This is due to the dishes only being spaced 60\,cm from one another, and these new feeds having no cages surrounding them which would otherwise limit this particular systematic effect \citep{Fagnoni_2021,Josaitis_2022}.

Whether cross-coupling spectra undergo cable reflections or not depends on where in the signal chain these effects occur, but reflected cross-coupling is generally considered to be of a low enough order that it is expected to be below the noise level. 
In \cite{Aguirre+22}, the cross-coupling model is written as
\begin{equation}\label{eq:CC}
    V_{apbq,t}^{cc} = V_{apap,t}^{\rm{refl}}\left( \sum\limits_{j}^{\rm{N}}{\rm{A}}_{apbq}^{d,j}\exp(i2\pi\nu\tau_{apbq}^{d,j}+i\phi_{apbq}^{d,j})\right)_{d\ni t} ,
\end{equation}

where $\rm{N}$ is the number of cross-coupling peaks present in the visibility, and the free parameters are the amplitude ($\rm{A}$), delay ($\tau$), and phase ($\phi$). $V_{apbq,t}^{cc}$ is the cross-coupling spectrum which corrupts the true data, and $V_{apap,t}^{\rm{refl}}$ is the copied autocorrelation which may possibly already contain cable reflection systematics. As with the reflections, the indices $p$ and $q$ refer to the feeds of antennas $a$ and $b$, respectively, and $t$ the integration time. Since cross-coupling is expected to be time-stable over a night of observation, $d$ refers to a particular day. Eq. \ref{eq:CC} only includes the cross-coupling contribution from antenna $a$, which is located at positive delays. As with all systematics analysis in this work, we only focus on those at $\tau>0$.

This coupling visibility is added to the cross-correlation as
\begin{equation}\label{eq:CCApply}
    \tilde{V}_{apbq,t}^{\rm{corrupt}} = V_{apbq,t}^{\rm{refl}} + \begin{cases}
            V_{apbq,t}^{cc} & a\neq b\\
            0 & a = b 
            \end{cases}.
\end{equation}
$V_{apbq,t}^{\rm{refl}}$ is the coupling-free cross-correlation, which possibly contains reflection systematics. $V_{apbq,t}^{cc}$ is only added in the case of antenna $a\neq b$, i.e. an antenna cannot couple with itself. 

The phases of the cross-coupling features are arbitrarily set by the coupling coefficient, $\epsilon$, but are time-stable. The cross-coupling delays are also unchanging, with only the amplitudes changing as a function of the autocorrelation amplitudes. These, in turn, vary as a function of the beam crossing, which is around one hour for HERA \citep{Kern+19}. For the 21.4 second cadence of HERA's redundant time averaging \citep[which occurs prior to systematic removal;][]{Abdurashidova_2023}{}{}, cross-coupling can essentially be seen as being time-stable overall for the purposes of modelling and mitigation.

\subsection{Current Mitigation Methods}

Following their modelling, \cite{Kern+19} calibrate simulated data in order to recover the underlying signal. For the cable reflections, the coupling coefficient (Eq. \ref{eq:epsilon}) of each individual reflection peak is solved for. 

The delay of the peak of the reflection ($\tau_{\rm peak}$) is estimated in delay space via quadratic interpolation, where an initial estimate of the peak is found by comparing the visibility amplitude at a particular delay to the amplitudes at adjacent delays. The delay corresponding to the greatest power is taken as the location of the peak. The amplitude of the reflection peak ($\rm{A}$) is then estimated by the ratio of the visibility at $\tau_{\rm peak}$ and the visibility at $\tau = 0$ (the foreground peak):
\begin{equation}
    \rm{A}_{11} = \frac{|\tilde{V}(\tau = \tau_{\rm peak})|}{|\tilde{V}(\tau = 0)|} .
\end{equation}
This reflection is then isolated by zeroing visibility modes on either side of this peak in order to obtain the filtered visibility $V_{\rm{filt}}$. The phase is estimated by adjusting $\phi \in [0,2\pi)$ in order to minimise the quantity $|V_{\rm{filt}} - \rm{A}\exp(i2\pi\nu\tau + i\phi)|^2$. This provides estimates of all three parameters: $\rm{A}$, $\tau$, and $\phi$. These estimates are then adjusted with a nonlinear optimisation method where they are continuously perturbed, and the reflections are iteratively calibrated out of the data using these perturbed values until the reflection peaks are minimised as much as possible. 

\cite{Kern+19} remove cross-coupling with a singular value decomposition (SVD) approach. In delay space, an SVD is applied to the cross-correlation visibilities after downweighting the foregrounds in order to maximise the contributions from the coupling systematics.

In essence, this method acts as a decomposition of the matrix $\mathbf{\tilde{V}}=\mathbf{TSD^\dag}$. The matrix $\mathbf{T}$ contains the basis vectors across time, $\mathbf{D}$ the basis vectors across frequency, and $\mathbf{S}$ the eigenvalues.

The first few dominant modes following the SVD are chosen to describe the coupling. The eigenvalues, $\mathbf{S}$, along with the corresponding eigenmodes, $\mathbf{T}$ and $\mathbf{D}$, are used to form a cross-coupling model. This model is then subtracted from the data in order to recover the underlying visibility.

To improve the resistance to signal loss, the $\mathbf{T}$ modes are low-pass filtered, with a Gaussian process regression filter providing the best results. This ensures that only low-fringe-rate modes (where the coupling systematic is predominantly located) undergo subtraction, while the signal itself dominates high fringe rates. 

This could not be immediately replicated with our method as we do not form similar modes in our model. We instead form model visibilities which are compared to the mock data. The closest approximation would be, for multiple integrations, to filter our model cross-coupling-only visibilities across time, and using these to subtract this systematic from the mock data. This would possibly eliminate high-fringe-rate structure in our coupling model, as it does in \citet{Kern+19}'s method. However, our intent is to assess the viability of forward-modelling the systematics on a per-antenna, per-time-integration basis, without any additional filtering, SVD approaches, etc.

In the single cable reflection regime (where there is only one reflection peak, or peaks are clearly spaced out in delay), the algorithm of \cite{Kern+19} is capable of effectively suppressing the reflection with almost no signal loss. In this single-reflection regime, approximately 2\% of the foreground and EoR signal is lost in reflection calibration. The signal loss metric is a ratio between the recovered power spectrum and the true power spectrum as a function of delay. Wherever this ratio is $<1$ is considered a loss. This metric is calculated using an ensemble average of many realisations of a sky signal, and this value of 2\% is within the ${\rm{N}}^{-\frac{1}{2}}$ sample variance of the average. In other words, this calibration does not significantly attenuate the true sky signal.

In the multi-reflection regime (when there are multiple reflections present, but they are widely distributed in delay space and do not overlap), the algorithm can once again suppress the reflections with some signal loss in the autocorrelations. When this calibration is applied to the cross-correlations, there is essentially no signal loss (determined by plots of the signal loss metric as a function of delay). 

However, the reflections cannot be completely calibrated out when there is a significant number of almost overlapping peaks, leaving non-negligible residual reflections. In the autocorrelation power spectrum, the reflections can only be suppressed by around four magnitudes out of the required $\sim10$ (for \citet{Kern+19}'s example). When these results are applied to the corresponding cross-correlations, this corresponds to a suppression of around four orders of magnitude, where around six is needed. This is due to the confusion between individual systematics, which makes it difficult to discern where individual peaks start and end, which in turn makes them difficult to model and calibrate out. Despite the lack of complete reflection mitigation, there are again no significant levels of signal loss in the cross-correlations.

For cross-coupling, shorter cross-correlation baselines show some resistance to systematics removal. For the shortest baseline (15 m east-west in \cite{Kern+19}'s example), the cross-coupling can be mitigated, but the highest amplitude peaks leave approximately 5\% too much power. However, there is no significant signal loss. There is better performance in longer baselines (29 m east-west), as cross-coupling is more separated from the EoR signal in fringe-rate space. While the systematic can again not be removed completely, especially the strongest features, there is about four orders of magnitude improvement over the short baseline. Overall, any signal loss produced by \cite{Kern+19}'s cross-coupling removal is on the order of $\sim1\%$.

These methods were applied to early HERA Phase I data \citep{DeBoer_2017} in \cite{Kern+20}. This consisted of a single 8 hour night's worth of data, recorded with 46 antennas. Reflections in individual antennas arose from a 150\:m cable connected to a post-amplifier module, followed by a 20\:m cable connected to a digitiser. This data was also subject to the coupling systematics described in Section \ref{section:Systematics}. A ``systematic tail'' was also observed, which was attributed to possible subreflections. In addition to cross-coupling, \cite{Kern+20} were able to mitigate these systematics down to the noise floor in the power spectrum. 

\cite{Abdurashidova_2023} provides lowered upper limits on the EoR power spectrum using 94 nights' worth of Phase I observations, and again employs these mitigation methods for the reflection and cross-coupling systematics, albeit with a number of changes to the application. In particular, it was found that the structure of the cross-coupling changed as the array grew, while \cite{Kern+19}'s method assumes stability. To account for this, the cross-coupling and reflection calibrations were carried out on a per-epoch basis, where an epoch is a period of observing runs uninterrupted by changes to the array, or malfunctions, for example. Additionally, the setup of the calibrations were changed to better suit the newer data. For example, the number of terms used in the reflection fitting were increased to account for newer and longer cables, and the number of delay modes used in the cross-coupling modelling were increased in order to improve the residuals.

\section{Simulated Data}\label{section:Data}
This work uses the HERA Validation simulations \citep{Aguirre+22}, which include simulated point source and diffuse foregrounds, as well as simulated EoR signals. This dataset was used to validate the software pipeline used in \cite{Abdurashidova_2022b}. We added systematics and instrumental bandpass gains to uncorrupted datasets with \textsc{hera\_sim}\footnote{\url{https://github.com/HERA-Team/hera_sim}}. The bandpass gains are randomly generated on a per-antenna basis, and as a function of frequency. Starting with the default HERA H1C bandpass in \textsc{hera\_sim}, randomised white noise multiplied by a spread factor (10\% being the default) is then added. This spread factor is the standard deviation of the randomised gains from antenna to antenna, i.e. across all antennas in the array, the generated bandpass gains vary by up to a factor of 0.1.

These bandpass gains were not calibrated out, as we intend to follow the general approach of \cite{Kern+19} and \cite{Kern+20} as closely as possible. The methods presented in \cite{Kern+19} make no assumption on whether instrumental bandpass has been corrected for, and \cite{Kern+20} apply these methods to observational HERA data which contained bandpass. The authors do state, however, that bandpass calibration could lead to better defined systematics in delay space, which would in turn improve their modelling and mitigation. Our bandpass is smooth as a function of frequency, and so should not adversely affect modelling to any significant degree.

Throughout this work, then, the ``true'' visibilities and power spectra we intend to recover contain foregrounds, signal, and bandpass.

The Validation simulations consist of 1024 frequency channels covering 100--200 MHz. Only 512 frequency channels out of the total 1024 are used in this work, to account for any possible trimming of observational data due to RFI, for example, and to avoid requiring an excessive number of model parameters. Maintaining 512 channels should ensure that there are more data points than parameters in our model. In total, our foreground and systematics models amount to $\sim100$ parameters. A detailed discussion is given in Section \ref{section:Modelling}. Ideally, one would prefer even more observational data points in order to help constrain these parameters, but 512 is a fair compromise between modelling performance and real-world considerations.

In this work, true foreground and EoR visibilities are taken from the validation data to construct the corresponding models, as well as our mock data which consists of the total foreground + EoR visibility contaminated by systematics and noise. Our noise estimates are also formed with this data.

Using \textsc{hera\_sim}, similar parameters to those in \cite{Aguirre+22} were used to generate the systematics in our data. \cite{Aguirre+22} simulated systematics matching those seen in the analysis of \cite{Kern+20}.

With reference to Eq. \ref{eq:CableReflectionGain}, there are two high-amplitude cable reflections located at delays ($\tau$) of 200\,ns and 1200\,ns, with relative amplitudes ($\rm{A}$) of $3\times10^{-3}$ and $8\times10^{-4}$, respectively. In the simulations, the delays are randomised to within 10\,ns and 30\,ns, respectively, and the amplitudes are randomised to within 1\%. The phases ($\phi$) are randomised in the range $[-\pi,\pi)$. 

Twenty cable subreflections are added, again using Eq. \ref{eq:CableReflectionGain}. Initially, they are evenly spaced within the delay range 200--1000\,ns, but are then randomised to within 30\,ns. Relative amplitudes ranged from $10^{-3}$ to $10^{-4}$ with a randomisation within 1\%. Phases are again randomised within $[-\pi,\pi)$. 

Reflections appear at both positive and negative delays. They are symmetric in delay for autocorrelations, as the autocorrelations themselves are fully real and this systematic couples antennas with themselves. They are not necessarily symmetric in the cross-correlations, as these visibilities are formed from two antennas whose specific signal chain characteristics might differ, for example slightly different cable lengths or subreflection behaviour. We do, however, only focus the analysis on positive delays for simplicity.

Finally, in keeping with Eq. \ref{eq:CC}, ten cross-coupling spectra are added uniformly between delays ($\tau$) of 900-1300\,ns, thereafter randomised within 20 ns. Relative amplitudes ($\rm{A}$) range from $10^{-4}$ to $10^{-6}$, randomised to within 0.01\% of the initial value. Phases ($\phi$) are randomised in the same manner as before. Again, we only focus on positive delays. As it was observed and mitigated in the cross-correlations of HERA Phase I data \citep{Kern+20}, however, cross-coupling was not symmetric, with there being clear asymmetries between the positive and negative delay coupling peaks. It was determined that antenna $i$ contributed to the structure at negative delays in the cross-correlations $V_{ij}$, while antenna $j$ contributed to positive delays. The delays were a function of the distance between the leaking connection point and the antenna, and the amplitudes a function of how much a surrounding antenna's signal was being broadcast, as well as the distance it had to travel to the antenna in question.

All these systematics are assumed to be time stable. The work here focuses on the shortest east-west baseline ($\sim 14.7$\,m) which is more susceptible to EoR signal loss when mitigating systematics than longer baselines. This is due to the overlap between the EoR and cross-coupling systematics in fringe-rate space. Cross-coupling occupies lower fringe-rates as it varies slowly, and overwhelms a portion of the EoR. Shorter baselines suffer most from this, as the EoR occupies lower fringe-rates in this regime. As the baseline length increases, the EoR moves towards higher fringe-rates and away from this systematic (in turn making systematic mitigation without signal loss easier). Moving from east-west baselines to north-south complicates this, however, as the EoR moves back towards lower fringe-rates occupied by cross-coupling. For these north-south baselines, cross-coupling occupies all of the same fringe-rate modes as the EoR, possibly making mitigation of this systematic impractical without any significant signal loss \citep{Kern+19}.

\section{Modelling}\label{section:Modelling}
This section details the modelling of the foregrounds and EoR signals, as well as the setup of the Bayesian model estimation problem and model\footnote{\url{https://github.com/GeoffMurphy/HMCSystematicsSampler/}}.

\subsection{Foreground Model}\label{subsection:FG_model}

In order to avoid having to model observed foregrounds to high precision, HERA uses a foreground-avoidance technique in its observations \citep{Kerrigan_2018,Morales_2018}, 
which aims to retain only the Fourier modes outside of the foreground wedge, i.e. regions where the foregrounds are less dominant. This foreground wedge is typically defined in 2D Fourier space -- the plane formed by $k_\parallel$ and $k_\bot$, where the former is dependent on the spectral resolution and bandwidth of the array, and the latter on the baseline distribution. However, in this work, an assumed foreground model is required, both to construct the visibilities, and to simulate the systematics which produce copies of these foregrounds at higher delays.

\begin{figure}
    \centering
    \includegraphics[width=1\columnwidth]{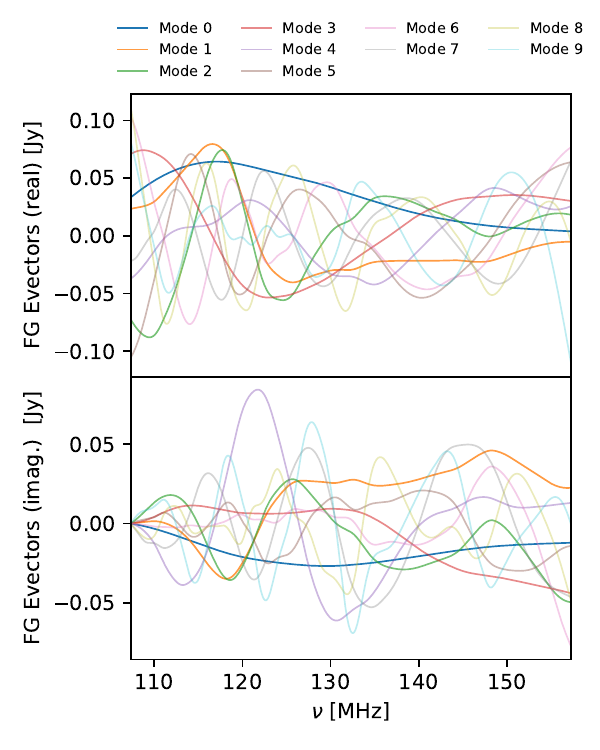}
    \caption{An example of ten cross-correlation foreground eigenvectors for a 14.7\,m east-west baseline, with the top being the real components, and the bottom the imaginary. These ten modes are the eigenvectors corresponding to the largest eigenvalues in the decomposition, with ``Mode 0'' corresponding to eigenvalue 0 in Fig. \ref{fig:Cross_Evals}, ``Mode 1'' to eigenvalue 1, etc. Furthermore, they have the same units as the foregrounds in frequency space. The autocorrelation foregrounds exhibit similar eigenvector characteristics, albeit with only the real components being non-zero.} 
    \label{fig:Cross_Evecs}
\end{figure}

\begin{figure}
    \centering
    \includegraphics[width=1\columnwidth]{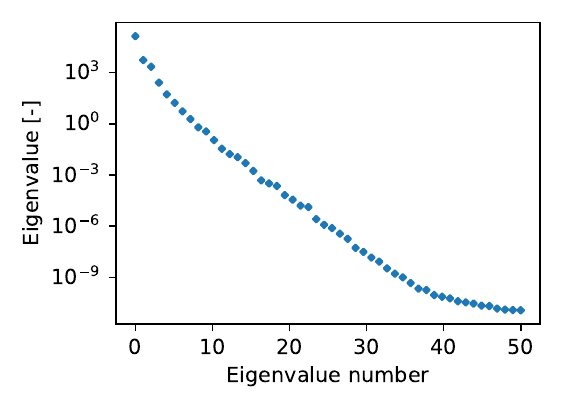}
    \caption{The first fifty cross-correlation foreground eigenvalues found through the eigenvalue decomposition. These are unitless, as they are factors multiplied with their corresponding foreground eigenvector, which subsequently forms a model foreground in units of Janskys.} 
    \label{fig:Cross_Evals}
\end{figure}

\begin{figure}
    \centering
    \includegraphics[width=1\columnwidth]{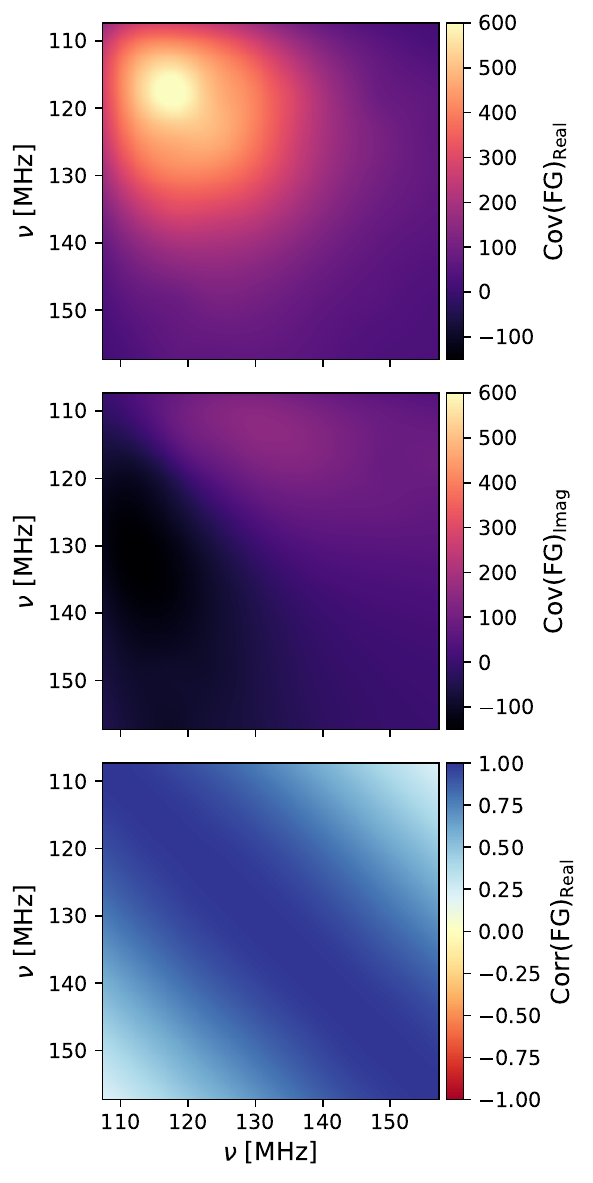}
    \caption{The covariance and correlation matrices of the HERA Validation simulated foregrounds for a 14.7\,m east-west baseline in units of $\rm{Jy}^2$. The top plot is the real component of the covariance matrix, the middle the imaginary component, and the bottom the correlation matrix.}
    \label{fig:Cov_Matrix}
\end{figure}

The simplified model used here is constructed from simulated HERA Validation foregrounds, which consist of the GaLactic and Extragalactic All-sky Murchison Widefield Array (GLEAM) point-source catalogue \citep{Hurley+17}, with additional bright sources added, and a diffuse component based on the extended Global Sky Model \citep[eGSM,][]{Zheng+17, Kim+18}. It was sufficient in our case (focusing on the shortest baselines) to construct the foreground model from the diffuse component only. Modelling the point-source components as well would likely require a number of higher-order modes, increasing the number of model parameters. 

Using the available 17,280 times for a particular antenna or baseline from the HERA Validation simulations, a frequency-space foreground model is constructed by calculating a set of basis vectors from an eigenvalue decomposition of the foreground visibility frequency-frequency covariance matrix. With 512 frequency channels, the covariance of the (complex) $512\times17,280$ data matrix ($\textbf{X}$) is found with $\rm{Cov}(\textbf{X}) = \textbf{X}\textbf{X}^\dagger$. The eigenvectors and eigenvalues are then found using \textsc{numpy}'s linear algebra routine \textsc{eigh}. The eigenvectors corresponding to the largest eigenvalues are used in the model, with the amplitudes of these vectors being used as free parameters in the model. 

Fig. \ref{fig:Cross_Evecs} plots the first ten eigenvectors for the cross-coupling foreground model. Bandpass gains were applied to the foregrounds prior to the eigendecomposition (but were not calibrated for). The strongest modes model the most prominent features in the foregrounds, which in this case is the smoothest structures across frequency. Fig. \ref{fig:Cross_Evals} shows the first fifty eigenvalues from the decomposition. The behaviour is as expected, with there being a decrease in eigenvalue amplitudes towards higher modes. Fig. \ref{fig:Cov_Matrix} shows the covariance and correlation matrices for the simulated diffuse foregrounds from which our eigenmodes are constructed. Eight eigenvectors were required to sufficiently recover the foregrounds, for both the autocorrelations and cross-correlations.

\subsection{EoR Model}\label{subsection:EoR_Model}

\begin{figure*}
    \centering
    \includegraphics[width=1.8\columnwidth]{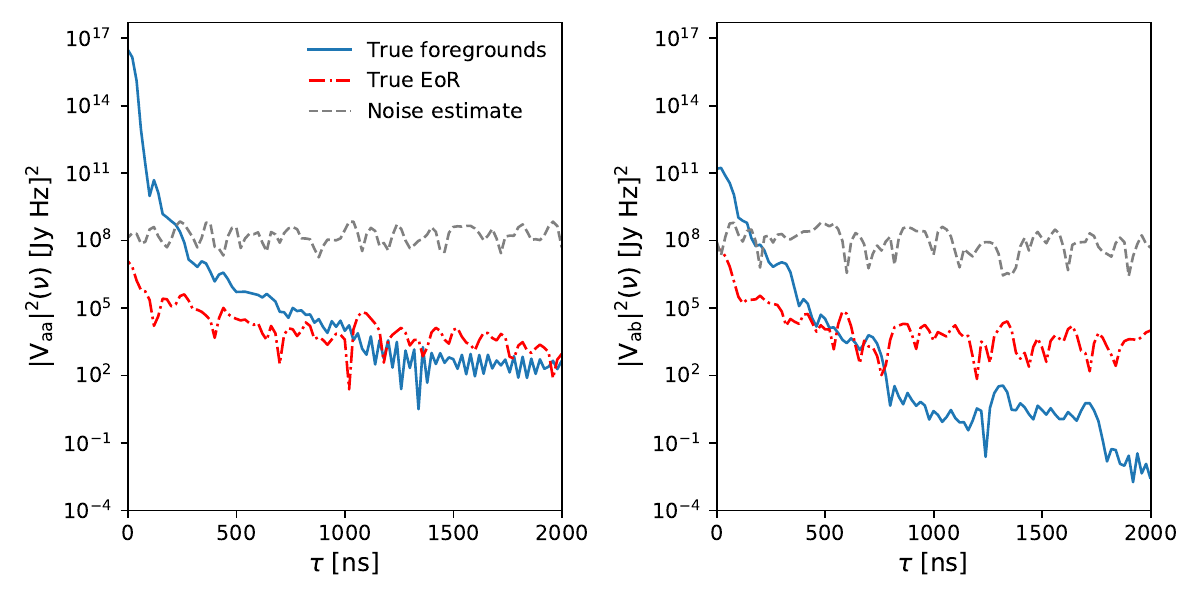}
    \caption{Left: The autocorrelation power spectrum of the noise estimate (grey dashed line) for a single time integration in comparison to the true EoR (red dot-dashed line) and true foreground (blue solid line) power spectra across the delay range of interest. Right: A comparison with the noise estimate in the cross-correlation regime, in this case for a 14.7 m east-west baseline. The noise estimate is derived from Eq. \ref{eq:LikelihoodStdDev}. }
    \label{fig:Noise_PS}
\end{figure*}

The model was also tested with and without an included signal component. Initially, it was thought that the approximately constant power across delay needed to be accounted for (see Fig. \ref{fig:Noise_PS}), but it was found that the results were comparable between both cases. This is due to the power of the systematics, which are much higher than the signal, so the lack of a signal component does not adversely affect modelling to any significant degree. The EoR model tested here is formed from the full-sky, full-bandwidth, and physically motivated mock-EoR simulated in the HERA Validation simulations \citep{Aguirre+22}, where it is modelled as a Gaussian random brightness temperature field. 

For a particular baseline and polarisation, an example EoR visibility is Fourier transformed, and $200$ complex Fourier modes are uniformly sampled from the delay range $(-2000,+2000)$\,ns, which contains the entirety of the foreground peak and all of the systematics considered in this work. Fewer modes would result in gaps in delay space. Our model simply fits each of the 200 Fourier modes individually with 200 amplitude parameters. This is discussed further in the context of priors in Section \ref{section:Priors}.

It was found that in realistic, noisy data, the recovered power spectra from the signal-free model was essentially equivalent to that found with a model containing a signal.

While the inclusion of a signal model slightly improved modelling in essentially noise-free cases (namely in the form of narrower uncertainty estimates in some cases), as it is implemented here with 200 Fourier modes, the run-times became prohibitive. This is due to a combination of the low noise, as well as the number of added parameters. This improvement also presented itself as better recovery of the underlying power spectra, but for reflection modelling this was a negligible gain in the very-low noise regime (i.e. it was comparable to the result of Fig. \ref{fig:chain_cross_sub_no_noise}). For cross-coupling mitigation with very-low noise, the recovery improved by approximately half an order of magnitude with the inclusion of a signal model. Overall, the benefits of the inclusion of this component were outweighed by the performance loss in negligible-noise scenarios, and its inclusion was not necessary in high noise regimes.

Nevertheless, realistic-noise cases are more relevant. Here, the inclusion of a signal component provides comparable results to a signal-free model. As such, we opt to exclude this component in our analysis, which also allows for faster sampling speed.

\subsection{The Systematics Model}

Combining the foreground and systematics terms in Eqs. \ref{eq:CableReflectionApply} and \ref{eq:CCApply}, respectively, we get the full systematics equation for our model:

\begin{equation}
    V^{\rm{corrupt}}_{apbq,t} = \sum\limits_{m=1}^{\rm{M}_{\rm{fg}}}\lambda_{abm} x_{abm} + \begin{cases}
    \tilde{g}_a\tilde{g}_b^* V^{}_{apbq,t'}  & a = b\\
    \tilde{g}_a\tilde{g}_b^* V^{}_{apbq,t'} + V_{apbq,t}^{\rm{cc}} & a \neq b
    \end{cases},
\end{equation}

where 

\begin{equation}\label{eq:Vcc_full}
    V_{apbq,t}^{\rm{cc}} = \tilde{g}_a\tilde{g}_b^* V^{}_{apap,t'}\left( \sum\limits_{j}^{\rm{N}_{CC}}{\rm{A}}_{apbq}^{d,j}\exp(i2\pi\nu\tau_{apbq}^{d,j}+i\phi_{apbq}^{d,j})\right)_{d\ni t}.
\end{equation}

Here, $x_{abm}$ are the foreground eigenvectors, and $\lambda_{abm}$ the corresponding free parameters for the foregrounds measured by antennas $a$ and $b$, with there being a total of $\rm{M}_{\rm{fg}} = 8$ such modes in our model. As described in Section \ref{subsection:CR}, $\tilde{g}_a$ and $\tilde{g}_b$ are our reflection gain terms (written in full in Eq. \ref{eq:CableReflectionGain}), which includes both high-amplitude cable reflections and subreflections (i.e. $\rm{M}=22$ in Eq. \ref{eq:CableReflectionGain} for the fiducial model). 

${\rm{A}}_{apbq}^{d,j}$ are the cross-coupling relative amplitudes described in detail in Section \ref{subsection:cross_coupling}. Along with the delay, $\tau_{apbq}^{d,j}$, and phase, $\phi_{apbq}^{d,j}$, parameters, a total of $\rm{N_{\rm{CC}}}=10$ such features are modelled, where only the contribution from one of the two antennas are considered in this work, i.e. the coupling features present at positive delays. We also assume the coupling systematics undergo cable reflections, which accounts for the $\overline{g}_a\overline{g}_b^*$ factors in Eq. \ref{eq:Vcc_full}. As in the previous paragraph, there are 22 such reflection/gain terms. These are, however, typically very weak, with relative amplitudes of $10^{-7}-10^{-10}$, and so are negligible, but are still present in our model as result of the order in which systematics are added, i.e. the model adds coupling systematics first, followed by reflections, so reflected coupling is a natural byproduct of this.

Since we only model positive delays, we neglect the third term for the $a\neq b$ case. This would consist of the visibility from the second antenna, $V^{\rm{uncal}}_{bqbq,t'}$, with distinct amplitudes, delays and phases for the coupling model. There would also be a leading factor of $\tilde{g}_b\tilde{g}_a^*$ denoting the cable reflection of the coupling-affected visibility.

\subsection{Noise}

Our thermal noise estimate is derived from the radiometer equation \citep[e.g.][]{Choudhuri+21} as
\begin{equation}\label{eq:LikelihoodStdDev}
    \pmb{\sigma}_{ij} = \sqrt{\frac{V_{ii}V_{jj}}{\Delta t \Delta \nu}},
\end{equation}
where $\Delta \nu = 97.7$ kHz is the frequency resolution of the simulated data, with an integration time of $\Delta t=10.73$\,seconds. $V_{ii}$ and $V_{jj}$ are the autocorrelation visibilities for the two antennas in a particular baseline when analysing a cross-correlation. 

Using the noise estimate of Eq.~\ref{eq:LikelihoodStdDev}, which is fully real, a complex noise visibility can be formed with
\begin{equation}\label{eq:noise_spectrum}
    \pmb{n}(\tau)_{ij} = \frac{1}{\sqrt{2}}\pmb{\sigma}_{ij}(\tau)\textbf{X}_1 + i\frac{1}{\sqrt{2}}\pmb{\sigma}_{ij}(\tau)\textbf{X}_2,
\end{equation}
where $\textbf{X}_1$ and $\textbf{X}_2$ $\sim N(\mu=0,\sigma=1)$ - random draws from a normal distribution with mean zero, and a standard deviation of 1. The random draws differ between the real and imaginary components. $\pmb{n}(\tau)_{ij}$ is an array with shape 512. This is Fourier transformed and squared in order to provide an estimate of the noise level in the power spectrum regime. Fig. \ref{fig:Noise_PS} provides examples of the autocorrelation and cross-correlation power spectra and their associated noise power spectra for a single 10.73 second time integration.

\subsection{The No-U-Turn Sampler}\label{section:Sampler}

Sampling is carried out with \textsc{Pymc3} \citep{Salvatier+16}, which allows models to be fit to data using a range of methods, for example Markov Chain Monte Carlo (MCMC), Hamiltonian Monte Carlo (HMC), or Gaussian processes. A self-tuning implementation of HMC \citep{Duane+87,Betancourt+17} named the No-U-Turn Sampler (NUTS) is the default and is used here \citep{Hoffman_2014}. This is instead of the more commonly used MCMC algorithms, which are less effective for models with parameter numbers on the order of tens or more due to the `curse of dimensionality' \citep{zeus}. NUTS also has the advantage of being self-tuning. The sampler automatically sets a number of more complex model hyperparameters, allowing for easier use in comparison to the HMC algorithm. For example, NUTS automatically adjusts the sampling step size, a hyperparameter to which HMC is very sensitive.

\textsc{pymc3} has built-in probability distributions, and automatically forms the likelihood function from the chosen distribution. The user, therefore, does not need to explicitly define their own likelihood functions. We use a normal distribution to evaluate the agreement between our model and the mock data, which is simply defined as
\begin{equation}
    \mathcal{L}(\pmb{x} \mid \pmb{d}, \pmb{\sigma}) = \frac{1}{\pmb{\sigma}\sqrt{2\pi}}\exp\left\{ -\frac{1}{2\pmb{\sigma}^2} \left (\pmb{m}(\pmb{x})-\pmb{d} \right )^2 \right\} ,
\end{equation}
where $\pmb{d}$ is the mock data, $\pmb{m}(\pmb{x})$ is the model as a function of the model parameters, and $\pmb{\sigma}$ our noise rms \citep{Salvatier+16}. The data, model, and noise are vectors of shape 512 (the number of frequency channels) for each baseline and time.

For the most part, only a few changes are made to the default settings of the sampler. \textsc{Pymc3}'s ADVI \citep[Automatic Differentiation Variational Inference,][]{Kucukelbir_2015} initialiser is used, with there being $10^6$ initialisation steps, although initialisation completes long before reaching this limit. Having too few initialisation steps can interrupt this process before the loss value converges, adversely affecting sampling performance. This initialiser sets up a scaling matrix for NUTS, which approximates the posterior distribution and informs the size of the steps taken \citep{Salvatier+16}. 

Each chain in the sampler is then tuned for 5,000 steps, and sampled for a total of 10,000 steps with a target acceptance rate of 0.8. When sampling, the step size used in the model is adjusted in order to achieve the given target acceptance rate. A larger target acceptance can help with sampling of complex posteriors (e.g. those which deviate from a Gaussian in shape), but can also result in slower sampling \citep{Betancourt+17}. 

Two chains are sampled, and their results are combined. A chain is essentially a single sampling run, and \textsc{Pymc3} allows multiple chains to be run simultaneously. Sampling with multiple chains provides not only more samples, but also acts as a diagnostic tool. For example, if the posterior distribution is problematic (such as being multimodal), multiple chains are more likely to sample this effectively. A single chain might ``get stuck'' in a region of relatively high likelihood, and might never explore other regions. Sampling with multiple chains can help ameliorate this, as the individual chains are typically initialised in different regions of parameter space \citep{Betancourt+17}.

\subsection{Priors}\label{section:Priors}

In this section we outline our prior choice methodology. These are our best estimates for the model parameters before sampling. When generating the mock data, a level of randomisation was introduced in all systematics parameters. As such, it was assumed that we did not have perfect knowledge of the systematics' characteristics. Coupled with the noise, estimates were chosen to take this randomisation into account. For example, if a power spectrum was visually inspected, and a reflection systematic was estimated to be located at a delay of 200\,ns, the prior was set to be a normal distribution centred on this delay, with a standard deviation which takes into account the uncertainty of its true delay, the noise, and perhaps any confusion between closely neighbouring systematics. The sampler then evaluates these priors against the mock data in order to form the posterior distributions, which will, ideally, be centred on the true delay.

\begingroup

\begin{table*}
\begin{center}
\renewcommand{\arraystretch}{1.4} 
\begin{tabular}{ |c c|c|c| } 
 \hline
Parameter name (no. of parameters) & [unit] & Prior mean & Prior std. dev. \\[0.1cm] 
 \Xhline{2\arrayrulewidth}
\multicolumn{4}{|c|}{Autocorrelation foregrounds \& EoR}\\
 \hline
        Foreground eigenvector amplitudes (8) & -- & $10^{2}-10^{-3}$ & $5\%$ \\
 \hline
\multicolumn{4}{|c|}{Cross-correlation foregrounds \& EoR}\\
\hline
        Foreground eigenvector amplitudes (8) & -- & $10^{-1}-10^{-3}$ & $5\%$ \\
\hline
\multicolumn{4}{|c|}{Systematics}\\
\hline
        Cable reflection amplitudes (2) & -- & $\{5\times10^{-3}, 8\times10^{-4} \}$ & $1\%$ \\
        Cable reflection delays (2) & [ns] & $\{200, 1300 \}$ & $5$ \\
        Cable reflection phases (2) & [rad] & $0$ & $1$ \\[0.1cm] 
\hline
        Subreflection amplitudes (20) & -- & $10^{-3} - 10^{-4}$ & $10\%$ \\
        Subreflection delays (20) & [ns] & $200 - 1000$ & $1$ \\
        Subreflection phases (20) & [rad] & $0$ & $1$  \\[0.1cm] 
\hline
        Cross-coupling amplitudes (10) & -- & $10^{-4} - 10^{-6}$ & $2\%$ \\ 
        Cross-coupling delays (10) & [ns] & $900 - 1300$ & $5$ \\ 
        Cross-coupling phases (10) & [rad] & $0$ & $1$ \\[0.1cm] 
\hline
\end{tabular}
\end{center}
\caption{A description of all the priors used in the model, as well as how many of each parameter there are, and their units. All priors are normal distributions. For brevity, only the ranges and orders of magnitudes of some parameters are given. This is to provide a broad idea of how the prior widths compare to the means, and is due to fact that an exhaustive list would likely not be informative. The foreground amplitude prior means were derived from a vector projection of the model eigenvectors. Systematic amplitudes are measured relative to the foreground peak, and delays are in units of seconds. All phase priors are $\mathcal{N}(\mu=0,\sigma=1)$, which covers the $[-\pi,\pi)$ range. Subreflection amplitude prior widths are wider than other systematics as it is not immediately evident through visual inspection how much of the amplitude at a given delay is from the subreflection in question, neighbouring subreflections, or the foreground and EoR components. Note that the prior means are not necessarily identical to the parameters used to generate the systematics. When generated, randomisation is introduced, so the priors can be best initial guesses with this in mind.}
\label{table:Priors}
\end{table*}

\endgroup

Table \ref{table:Priors} describes the priors used in the sampler. When choosing priors, the amplitude and delay of a particular systematic can be fairly easily estimated. By Fourier transforming an observational visibility to delay space, the amplitude of the systematic peak can be compared to the foreground peak, and a relative amplitude can be inferred. Similar can be said of the delay of the systematic, where one can simply estimate the delay of the systematic peak visually, or from knowledge of the array's geometry. In observational data, this is naturally complicated by noise, data cuts, etc. 

Phases cannot be quickly and easily estimated by visually analysing the data, however. In delay-space, the phases act to broaden or narrow the systematics peaks. One would, in theory, need to apply some fitting or optimisation technique to get an initial guess of the phases. However, it is sufficient, for our purposes, to simply set normal priors on the phases with a mean of zero and a variance of $\BigOSI{1}{}$. This provides acceptable coverage of the [$-\pi$,$\pi$) range. Uniform priors across this range would likely perform similarly (i.e. drawing from a flat distribution across $-\pi$ to $\pi$), but priors such as this produce sharp decreases in posterior space at values corresponding to the edges of the prior. Sampling generally performs better when the posterior space is `smooth' and can be freely explored without strong restrictions. While uniform phase priors are generally sufficient in models with vastly fewer parameters, the high dimensionality of the model used here necessitates that we attempt to avoid decisions which would otherwise adversely affect performance.

In order to obtain reasonable means for the foreground eigenvector amplitude priors, the foreground eigenvectors, $\textbf{x}_{\rm{FG}}$, were projected onto one of the original simulated foregrounds, $V_{\rm{FG, true}}$. Specifically, this was $\lambda_{\rm{proj}} = {\textbf{x}_{\rm{FG}}} \cdot {V_{\rm{FG, true}}}$, as discussed in Section \ref{subsection:FG_model}. The resulting amplitudes, $\lambda_{\rm{proj}}$, were used as the prior means in the sampling. 

Furthermore, fair knowledge of the foregrounds were assumed (in terms of prior specificity). The amplitudes of the prior means for the autocorrelation foregrounds ranged from $\mathcal{O}(10^{2}) - \mathcal{O}(10^{-3})$. Normal prior standard deviation widths of $5\%$ the amplitudes were chosen. Cross-correlations are lower in power, with the absolute values of these amplitudes ranging from $\mathcal{O}(10^{-1}) - \mathcal{O}(10^{-3})$. The prior standard deviations were reduced in order to maintain a relative width of $5\%$.

For the majority of systematic parameters, estimates of the delay and amplitude were made by visually inspecting the mock data, and prior widths reasonably wide enough to account for randomisation, noise, and confusion were chosen. In actual observations, however, knowledge of the cable lengths used, and the separation of the antennas could be used to estimate delays, but amplitudes would likely still need to be inferred from the data itself.

Given the number of parameters, however, priors for subreflections needed to be somewhat specific in order to minimise the effect on performance. A prior width of order 1\,ns was used for the delays. This also helps to mitigate against confusion to an extent, as the subreflection peaks are typically very closely spaced. Limiting the parameter space forces a fit on a particular subreflection peak.

Confusion typically results in the sampler struggling to constrain the contributions from each individual subreflection. When two or more systematic features are situated near to one another in delay, the resulting overall power can be attributed to multiple combinations of amplitude, delay, and phase for these subreflections. This can be ameliorated with better and more specific priors, but confusion can also make the initial prior selection difficult for the same reason.

Therefore, as much effort as possible was put into identifying the delays of the individual peaks before sampling, so that the prior width could be as specific as possible. Unfortunately, this means that if the initial guesses are poor or wrong, the sampler will struggle to find acceptable parameters to describe the mock data. Nevertheless, the priors chosen here produced reasonable results.

Lastly, for the systematics in question, our model consists of 66 free parameters for the reflection systematics, and 30 for cross-coupling. When implementing the reflection calibrator of \cite{Kern+19}, one supplies the upper and lower delay bounds of each individual reflection peak, along with the number of iterations desired for its suppression. The coupling calibration requires the upper and lower bound of the entire coupling shelf, and the number of SVD modes to be used. For the mock data in this work, we opted for 15 SVD modes. 

\subsection{Systematics Subtraction}

Following sampling, the parameter results from the model are used to remove the systematics from the mock data with the intention of recovering the true power spectrum.

For high-amplitude reflections and subreflections, this involves forming the gain terms (Eq. \ref{eq:CableReflectionGain}) from the sampler results for the amplitudes, delays, and phases. To recover the reflection-free visibility, the inverse of Eq. \ref{eq:CableReflectionApply} is applied to the mock data, i.e.
\begin{equation}
    V_{apbq,t'}^{\rm{recovered}} = (\tilde{g}_b^*)^{-1} (\tilde{g}_a)^{-1} V_{apbq,t}^{\rm{refl}} .
\end{equation}
This is done with all 10,000 samples from the model. This produces 10,000 recovered visibilities, from which the 95th percentile can be formed.

For cross-coupling, two models are formed from the results: the first containing only model foregrounds, and the second containing model foregrounds and cross-coupling. More specifically, using the foreground eigenvector amplitudes ($\lambda^{\rm{Model}}$) from the sampling results, a systematics-free visibility is formed. With the foreground eigenvectors ($\textbf{x}_{\rm{FG}}$) from Section \ref{subsection:FG_model}, this is formed with 

\begin{equation}\label{eq:modelfgeor}
    V_{\rm{FG}}^{\rm{Model}} = \lambda^{\rm{Model}}\textbf{x}_{\rm{FG}} .
\end{equation}
The second model visibility additionally contains the cross-coupling systematics, which are modelled with Eq. \ref{eq:CC}. 
\begin{equation}\label{eq:modelcccorrupt}
    V^{\rm{Model}}_{\rm{Corrupt}} = \lambda^{\rm{Model}}\textbf{x}_{\rm{FG}} +  V_{\rm{CC}}^{\rm{Model}}
\end{equation}
Taking the difference of Eqs. \ref{eq:modelfgeor} and \ref{eq:modelcccorrupt} produces a cross-coupling-only visibility,
\begin{equation}
    V_{\rm{CC\;Only}}^{\rm{Model}} = V^{\rm{Model}}_{\rm{Corrupt}} - V_{\rm{FG}}^{\rm{Model}}.
\end{equation}
This is then subtracted from the mock visibility in order to recover the underlying signal,
\begin{equation}
    V^{\rm{Recovered}} = V^{\rm{Mock}} - V_{\rm{CC\;Only}}^{\rm{Model}}.
\end{equation}
This is again done for all samples produced by the model, and percentile regions are formed from this collection of recovered signals.

\subsection{Signal Loss}\label{section:Signal Loss}

The following metric from \cite{Kern+19} was used to evaluate the level of foreground and EoR signal loss following systematic subtraction: 
\begin{equation}\label{eq:R3}
    R_3(\tau) = \frac{\langle P_3(\tau) \rangle}{\langle P_1(\tau)\rangle} .
\end{equation}
In line with the notation in \citep{Kern+19}, a subscript of 1 indicates a true signal, 2 indicates a systematics corrupted signal (not shown here), and 3 the recovered signal. Here, $P_1$ and $P_3$ are the delay power spectra of the true and recovered signals, respectively, and the ratio of ensemble averages over multiple times are taken. Signal loss is defined as when $R_3(\tau)<1$, suggesting either foreground and/or EoR signal has been lost in recovering the true visibility.

\section{Results}\label{section:Results}
We present systematic modelling results in autocorrelations and cross-correlations for varying noise levels: high-noise, corresponding to a single time integration, reduced noise, which assumes there has been some level of noise reduction prior to modelling, and very-low noise, where the noise power spectrum level is approximately $10\%$ that of the EoR. In this last case, plots of the signal loss metric as a function of delay are included, as we are able to mitigate the systematics down to, or close to, the true power spectrum. For the higher-noise cases, we can ideally only mitigate down to this noise level, so it is not as informative to include plots of the respective signal loss metrics. Where possible, we compare our results to that of \citet{Kern+19} when their methods are applied to the same data.

\subsection{Autocorrelation Systematics}\label{subsection:AutocorrelationSystematics}

\begin{figure*}
    \centering
    \includegraphics[width=1.7\columnwidth]{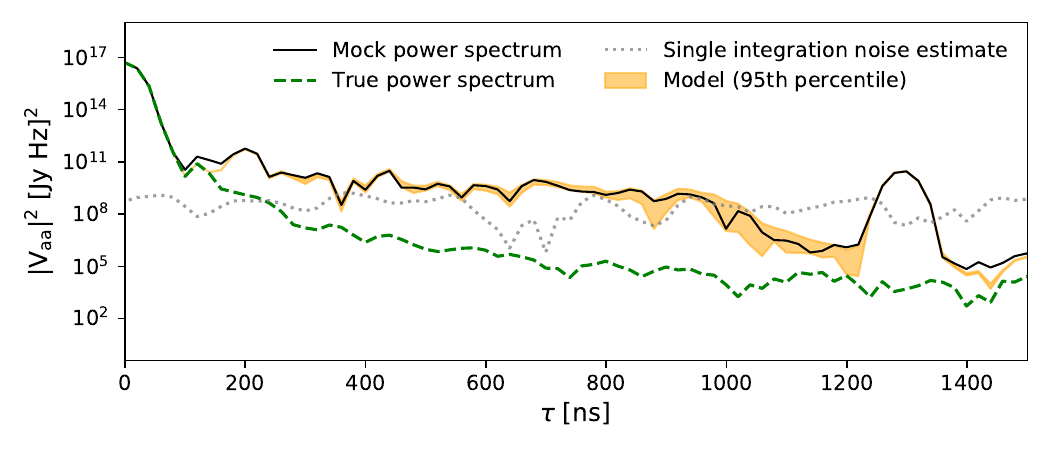}
    \caption{A comparison between the autocorrelation mock data (black solid line) and the model results (orange region), which sampled for two cable reflections, 20 subreflections, and 10 cross-coupling spectra in a single 10.73\,second integration. The mock power spectrum consists of the foreground, signal, and systematics, with the noise (grey dotted line) being taken into account when sampling. No constraints can be placed on features below the noise ($1000 \lesssim \tau \lesssim 1200$). The model is only centred on the mock power spectrum in this region because of the priors. Being below the noise level simply results in the model filling the prior space for these particular features. The model was run for 10,000 sampling steps, and the 95th percentile of these samples is shown here. The true power spectrum is denoted by the green dashed line, and contains only the foreground and EoR.}
    \label{fig:chain_auto}
\end{figure*}

Fig. \ref{fig:chain_auto} shows the sampling results for the full set of systematics (described in Section \ref{section:Data}) in a single 10.73\,second integration autocorrelation power spectrum using the priors and sampler settings described in Sections \ref{section:Sampler} and \ref{section:Priors}, respectively. For this plot, 10,000 power spectra are formed from the 10,000 samples, from which the $95\rm{th}$ percentile is calculated, denoted by the orange region.

Our priors are related to the noise-free mock data (black line), which is why the sampled power spectrum has features below the noise level. The reasoning is that the priors would be informed by the characteristics of the array, for example the length of cables for reflections, or the positions of antennas for cross-coupling. However, when sampling, we take the noise estimate into account by using Eq.\,\ref{eq:LikelihoodStdDev} as the standard deviation in the likelihood function. 

In the autocorrelation analysis, the full set of cable reflections, subreflections, and cross-coupling spectra is added to the mock data. While cross-coupling is much weaker in the autocorrelations than in the cross-correlations, the power is still non-negligible. However, there was little to no constraint on the cross-coupling parameters in this case, which was expected as the majority of the peaks are below the noise level.

Both the high-amplitude reflections and the majority of the subreflections are well constrained. Subreflections at delays higher than ${\sim800\,\rm{ns}}$ and cross-coupling spectra between ${\sim900-1300\,\rm{ns}}$ show a noticeable decrease in the level of constraint. This can be due to the confusion between systematics, but is mainly a result of the decreasing power of the systematics which brings them closer to the noise level.

\begin{figure*}
    \centering
    \includegraphics[width=1.8\columnwidth]{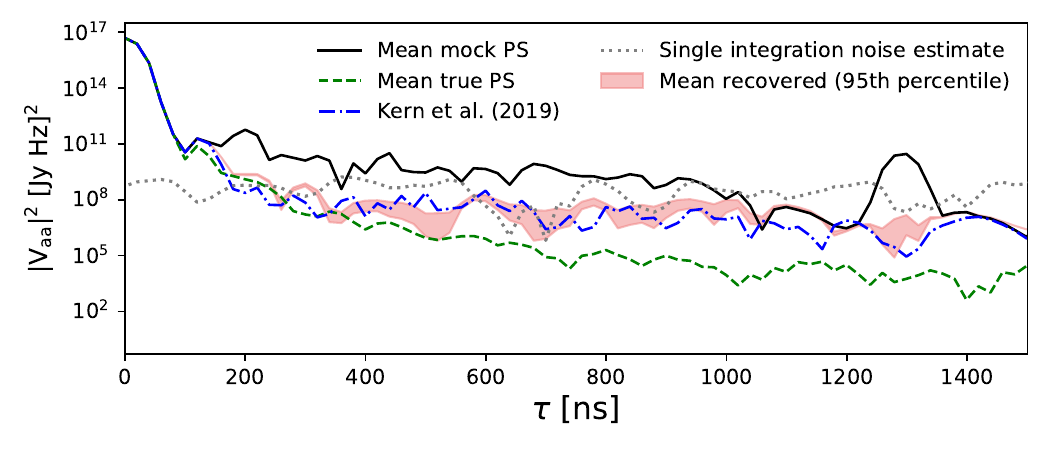}
    \caption{After mitigating the systematics in thirty time samples, each separated by 21.46 seconds, the incoherent time average of the recovered spectra are found, whereafter the percentile regions are formed, i.e. we go from $30\times10,000$ power spectra to $10,000$ mean spectra. The results of \citet{Kern+19} also correspond to the averaged recovered spectra when applied to the same data. The grey dotted line denotes the noise level for a single integration, and shows that noise can be mitigated following systematics removal.}
    \label{fig:Auto_Recovered_mean}
\end{figure*}

The model used above is applied to thirty visibilities each separated by 21.46 seconds. The systematics are mitigated from these visibilities, producing 10,000 recovered power spectra for each individual time sample, which are then incoherently averaged across time. From these 10,000 time-averaged recovered spectra, the 95th percentile is found, and is shown in Fig. \ref{fig:Auto_Recovered_mean}. Here, the thirty mock and true power spectra are time averaged as well, and are plotted against the noise level for a single integration. We are able to mitigate the systematics down to the single integration noise level, and thereafter incoherently average to further reduce the noise in our recovered spectra. Ideally, the objective in this case is for the band/confidence region to match the true power spectrum as closely as possible, where the true data only contains foregrounds, the EoR signal, and instrumental bandpass. 

The calibration results of \cite{Kern+19} when applied to the mock data are also included in Fig. \ref{fig:Auto_Recovered_mean}. Again, the calibration is applied to thirty time samples, and the power spectra are incoherently averaged across time. We see essentially the same behaviour, i.e. the systematics being mitigated down to the single-integration noise level, with there being further reduction in the noise after time averaging. 

As it is done in \cite{Kern+19}, we model the reflection systematics in the autocorrelation and subsequently apply the results to the corresponding cross-correlation. However, a cross-correlation visibility with a noise level consistent with the autocorrelations (i.e. derived from a single time integration, see Fig. \ref{fig:Noise_PS}) sees no useful results when the reflections are subtracted. When the results are applied, we effectively only recover the noisy cross-correlation, where the noise is on the order of $10^8\,\rm{Jy}^2\rm{Hz}^2$ in the power spectrum. The reflection systematics peak at approximately $10^5\,\rm{Jy}^2\rm{Hz}^2$. In this case, the noise is too strong to see any noticeable difference between the reflection corrupted and recovered power spectra, and so we do not present this particular result.

We omit the results for reflection mitigation in a reduced-noise case, where the noise is lower than in the above results, but still above the EoR level. As before, the systematics can be mitigated down to the noise level in the autocorrelation. In the cross-correlations, the noise is, again, well-above both the EoR power spectrum and the reflection systematics, so there is not much to be learned in this case.

\begin{figure*}
    \centering
    \includegraphics[width=1.8\columnwidth]{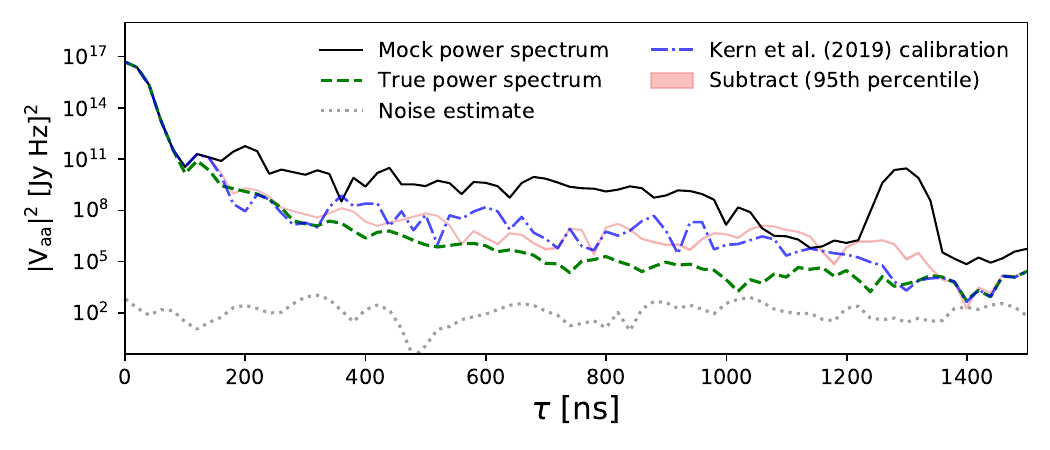}
    \caption{The recovered autocorrelation power spectrum when applied to data with very low noise. Here, we simply set the noise level to approximately 10\% the power of the EoR, rather than assume any particular integration time. As a result of the reduced noise, the width of the percentile bands becomes very narrow.}
    \label{fig:chain_auto_sub_no_noise}
\end{figure*}

If, instead, the model is run on visibilities with very low noise, where its power spectrum is $\sim10\%$ the EoR's level, we are then able to mitigate the systematics in the autocorrelations to a level closer to the true power spectrum (Fig. \ref{fig:chain_auto_sub_no_noise}). We mainly test this case to assess the signal loss properties of our model, and whether the mitigation removes primarily the systematics as it should, or if it suppresses the signal to too great a degree. There remains some residual systematics with powers of up to three orders of magnitude above the true power spectrum. This is primarily due to confusion between the systematics. \citet{Kern+19}'s method mainly suffers from this as well, where it is difficult to differentiate between the contributions from individual systematic features to the overall power. 

While we retain the same number of samples for Fig. \ref{fig:chain_auto_sub_no_noise}, as well as the formation of the 95th percentile, the bands do become much narrower as a result of the minimal noise, leading to tighter constraints on the systematics.

\begin{figure*}
    \centering
    \includegraphics[width=1.8\columnwidth]{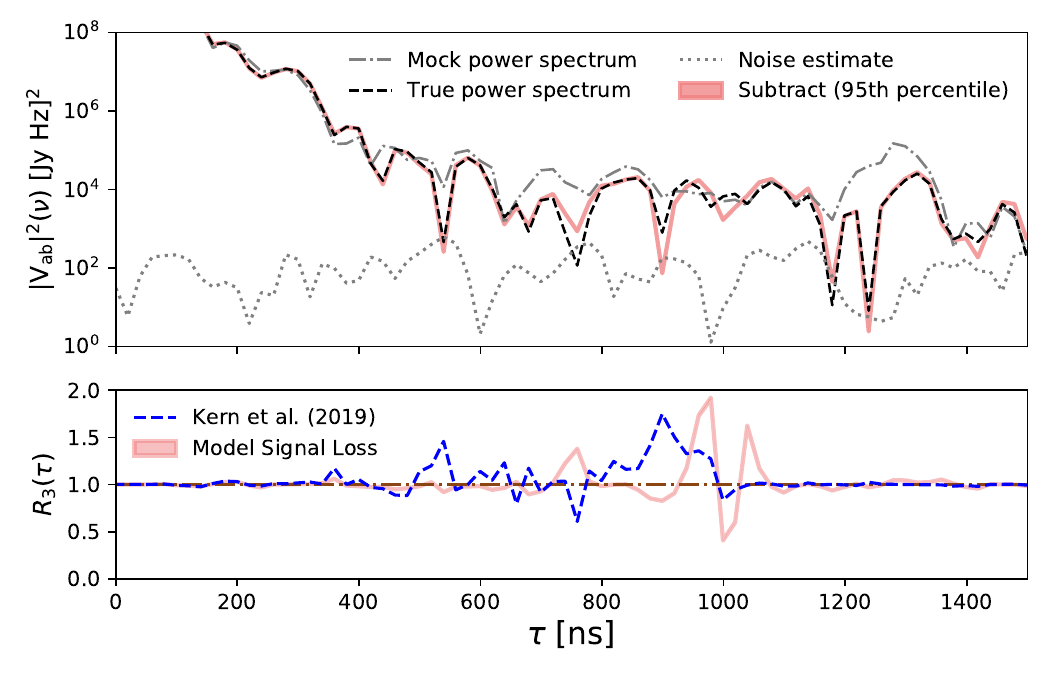}
    \caption{Top: The recovered cross-correlation power spectrum (light-red band) when the results of Fig. \ref{fig:chain_auto_sub_no_noise} are used to subtract the reflections from the 14.7\,m east-west cross-correlation mock data (grey dot-dashed line) in a low noise regime (denoted by the grey dotted line). We omit the results of \citet{Kern+19} for clarity and to prevent too many overlapping lines, but their model does perform similarly to ours, and their signal loss result is shown in the bottom plot. The objective, as before, is to recover the true power spectrum (black dashed line). The bottom plot, the signal loss metric, provides a comparison between our and \citet{Kern+19}'s results. The signal loss metric corresponding to our model is denoted by the light-red band, and that of \citet{Kern+19} by the blue dashed line. The $R_3(\tau)=1$ line is the brown, dot-dashed line, and denotes perfect recovery.}
    \label{fig:chain_cross_sub_no_noise}
\end{figure*}

By taking these results for the systematics parameters and applying them to the low-noise cross-correlation visibilities, we are mostly able to recover the true signal in Fig.\,\ref{fig:chain_cross_sub_no_noise} (top). With the reduction in the noise and the subsequent narrowing of the percentile bands, however, there are a number of delays where our results make somewhat poorer recovery, i.e. at $\sim750$\:ns and $\sim900-1100$\:ns. This is primarily a result of the performance in the corresponding autocorrelation (Fig. \ref{fig:chain_auto_sub_no_noise}) which carries over to the cross-correlation systematic mitigation. As detailed previously, the poor recovery in the autocorrelation is a result of confusion between the closely-spaced systematics. The clearest example of this is at $\sim900-1100$\:ns in Fig. \ref{fig:chain_auto_sub_no_noise}. Here, both confused subreflections and confused coupling systematics occupy the same delay range, leading to the largest amount of residual systematics, $\sim3$ orders of magnitude above the true power spectrum. This results in poorer recovery of the corresponding cross-correlation at similar delays, i.e. residual reflections or oversubtraction in Fig. \ref{fig:chain_cross_sub_no_noise}.

Where there are fewer classes of overlapping systematics, performance is improved, for example at $200-900$\:ns. While the model is still confusion-limited, the effects are not as pronounced when applied to the cross-correlation. Overall, however, this effect of confusion is ultimately a result of being unable to efficiently explore the parameter space. Normally one could choose wider, less-specific priors at the cost of slower sampling speed. However, sampling speeds already suffer as a result of the low noise, which produces prohibitive run-times. The alternative approach to mitigate confusion would be to choose extremely specific priors, but that would imply very good knowledge of the parameters and would defeat the purpose of sampling. Were the properties of the systematics known to such a degree, then a non-statistical mitigation approach would be sufficient.

Nevertheless, the model is capable of making a good recovery of reflection-corrupted cross-correlations. We opt not to include the result of \cite{Kern+19}'s method for clarity, but the performance is comparable in this case. The bottom plot of Fig.\,\ref{fig:chain_cross_sub_no_noise}, the signal loss metric, does provide a direct comparison between the two results, however. For the most part, our 95th percentile band adheres very closely to $R_3=1$, suggesting we are resilient to signal loss, much like \citet{Kern+19}. We see noticeably more residual systematics and signal loss at $900\lesssim\tau\lesssim1100$\:ns, a result of the overlapping reflections and coupling in the autocorrelation, but overall, the signal loss characteristics between the two results are very similar.

\subsection{Cross-coupling Systematics}\label{subsection:CrosscouplingSystematics}

This section focuses on cross-coupling in the cross-correlation power spectrum. While reflections are present in cross-correlations, they are typically very weak, and are instead modelled in the autocorrelations (as was done in Section\:\ref{subsection:AutocorrelationSystematics}). We opt to only model cross-coupling in this regime.

For a single time integration, the noise is approximately as strong as the highest amplitude cross-coupling peaks. With reference to Fig. \ref{fig:Noise_PS}, the noise power spectrum is of the order $10^8\,\rm{Jy}^2\,\rm{Hz}^2$. Cross-coupling amplitudes in the power spectrum are of this order of magnitude and lower. Essentially all of these systematic features, therefore, fall below the noise level, preventing any constraints from being placed.

\begin{figure*}
    \centering
    \includegraphics[width=1.7\columnwidth]{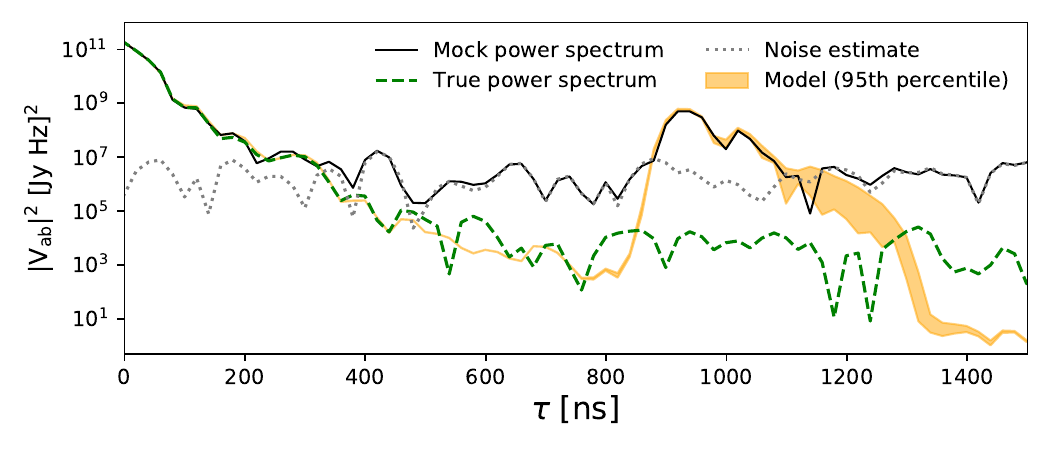}
    \caption{The model results for cross-coupling systematics in a reduced noise 14.7\,m east-west cross-correlation, where the integration time has been increased from 10.73\,s to 1073\,s, i.e. we have not incoherently averaged multiple power spectra here, but simply reduced the noise prior to modelling, for testing purposes. As before, the noise is plotted separately from the mock data for illustration purposes. The orange region is the 95th percentile of the 10,000 model power spectra samples. As discussed in Section \ref{subsection:EoR_Model}, we omit a signal component, which is why the power spectrum is unmodelled at delays away from the foreground and systematics peaks.}
    \label{fig:chain_cross_corr_tx100}
\end{figure*}

\begin{figure*}
    \centering
    \includegraphics[width=1.7\columnwidth]{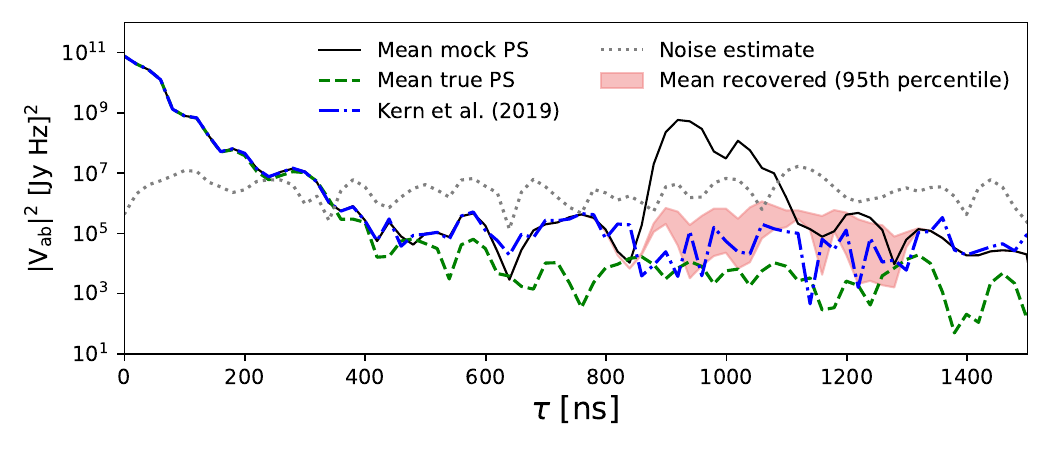}
    \caption{The power spectrum recovery when jointly modelling thirty time integrations separated by 21.46 seconds, recovering the underlying power spectra, and thereafter incoherently time averaging. Here, 10,000 samples are taken. The results of \citet{Kern+19}, the true power spectra, and the mock power spectra have all been averaged as well. The noise power spectrum corresponds to the level prior to incoherent averaging, and is the power at which sampling takes place.}
    \label{fig:Cross_Recovered_mean}
\end{figure*}

\begin{figure}
    \centering
    \includegraphics[width=0.9\columnwidth]{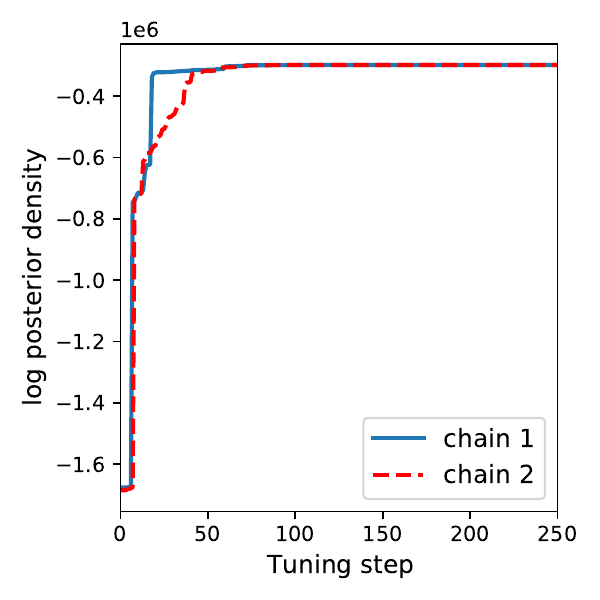}
    \caption{An example of the log posterior density during the tuning (burn-in) phase, corresponding to the sampling results of Fig. \ref{fig:Cross_Recovered_mean}. This plot shows the first 250 of 5,000 tuning steps for the two chains, demonstrating that the model tunes fairly quickly for this particular result. Sampling commences after this stage.}
    \label{fig:burn_in}
\end{figure}

\begin{figure*}
    \centering
    \includegraphics[width=1.7\columnwidth]{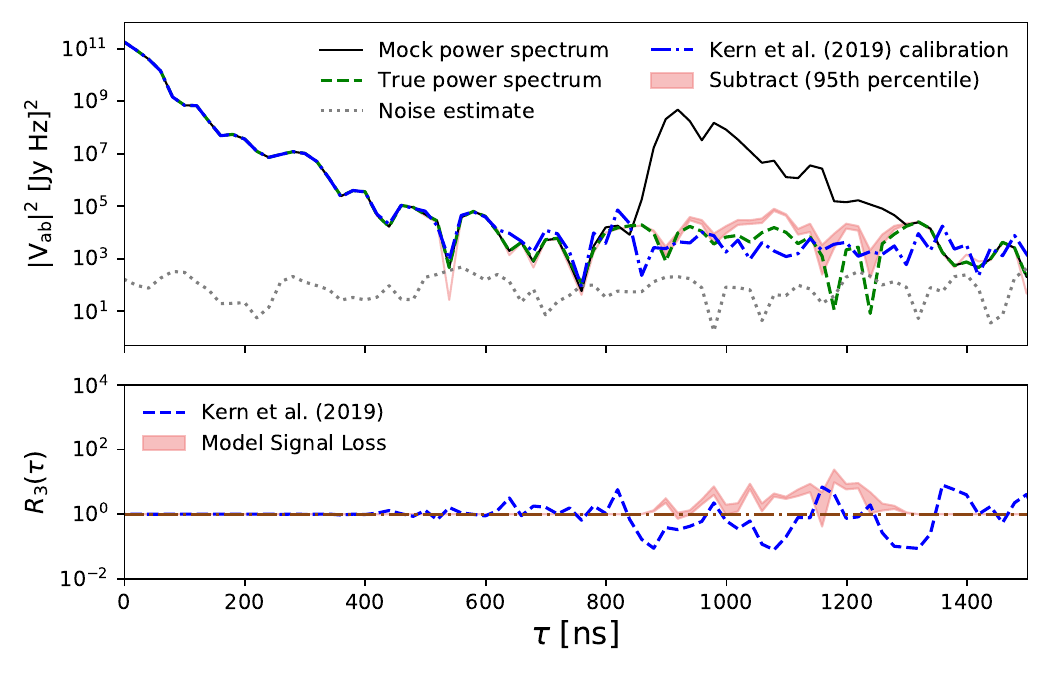}
    \caption{Top: The recovered cross-correlation power spectrum (red region) in the very-low-noise regime (denoted by the grey dotted line), where we assume the noise is below the signal. This is for a single mock spectrum, and no incoherent averaging is carried out for this result. This is again compared to the SVD method of \citet{Kern+19} (blue dot-dashed line), and the true, systematics-free power spectrum (green dashed line). Bottom: The corresponding signal loss results, with ours being the red region, and \citet{Kern+19}'s being the blue dashed line. The brown, dot-dashed line denotes $R_3(\tau)=1$ for perfect recovery.}
    \label{fig:chain_cross_corr_sub_no_noise}
\end{figure*}

This necessitates a reduction of the noise in order to constrain the cross-coupling systematics to any appreciable degree. Fig. \ref{fig:chain_cross_corr_tx100} shows the results when the integration time in Eq. \ref{eq:LikelihoodStdDev} is increased from 10.73\,seconds to 1073 seconds, i.e. assuming that noise has been reduced prior to systematics mitigation. While it is not so simple when considering observational data, we do not explore any complicating factors, such as the changing sky signal over time, and instead assume a reduced noise level so that we can test our model for this systematic. 

In this case, we are able to strongly constrain the highest amplitude cross-coupling peaks, with the level of constraint reducing as the noise level is approached (Fig. \ref{fig:chain_cross_corr_tx100}). The peaks below the noise level are not constrained at all, as is expected. Using these results, we can mitigate the cross-coupling systematics down to the noise level in a single integration, performing similarly to the method of \citet{Kern+19} when the first 15 SVD modes are used.

An integration time of $1073\:\rm{seconds}$ is considerably higher than the overall time averaging cadence in HERA analysis, which is $214$\,seconds. Furthermore, systematics calibration is carried out on data averaged to a $21.4$\,second cadence, with the remaining coherent time averaging occurring after the systematics are removed \citep{Abdurashidova_2023}. As such, this is not a representative example of noise reduction in the analysis pipeline, but nevertheless noise needs to be reduced in some manner in order to constrain cross-coupling systematics at all.

As with the reflection modelling, we apply the coupling model to thirty time integrations, mitigate this systematic, and incoherently time average all of the samples. However, following the same process as we did for the autocorrelations produced poor results. Namely, sampling each integration individually, recovering the power spectra, and thereafter averaging, did not produce a further reduction of the noise in the delay range of the systematics. This suggests that the model could constrain the coupling systematics only well enough to mitigate them down to the noise level ($\sim10^7\,\rm{Jy}^2\rm{Hz}^2$), but left residual systematics hidden by this noise. Attempting to average the recovered spectra made this additional structure evident, and the incoherently averaged power spectrum remained at $\sim10^7\,\rm{Jy}^2\rm{Hz}^2$ at delays of $900\lesssim\tau\lesssim1100\,\rm{ns}$, while delays dominated by noise saw a decrease in power.

If, instead, we jointly model all thirty time integrations simultaneously, then the results are improved. Here, we have thirty terms in our likelihood function, and the systematics priors and parameters are shared between all of them. In other words, all integrations have the same systematics solution solved for at once, rather than solving for each visibility individually, which is what was described in the paragraph above. Using the result from the joint model, we subtract the systematics from the thirty visibilities, and incoherently average the power spectra.

This is compared to the time averaged mock and true power spectra, as well as the recovered spectra from the \citet{Kern+19} calibration in Fig. \ref{fig:Cross_Recovered_mean}. We can mitigate the systematics down to the noise level for all thirty power spectra, and incoherently averaging produces a further reduction in the noise. Our result is comparable to the result of \citet{Kern+19}. Outside of differences in the finer features, the incoherently averaged recovered power spectra are at a power of $\sim10^5\,\rm{Jy}^2\rm{Hz}^2$, a reduction of around two orders of magnitude from the initial noise level. Our result has the benefit of an associated uncertainty, and for the most part there is agreement between the two results within this uncertainty.

We also provide an example of the tuning/burn-in phase for this particular result. \textsc{Pymc3} tracks the log posterior density, which is proportional to the product of the likelihood and prior of the model \citep{Lynch2007}. The aim is for the model to maximise the posterior density. As expected, there is an increase in this quantity as the model explores, and eventually settles in, the posterior space. There are 5,000 tuning steps for both chains, and following this the model draws a total of 10,000 samples, corresponding to the results of Fig.\:\ref{fig:Cross_Recovered_mean}.

The cross-coupling modelling performance was also tested in the very-low-noise regime, with the noise power spectrum being $\sim10\%$ that of the EoR. Using noise levels this low slows sampling severely, so it was opted not to reduce it further, and only a single integration was modelled.

We omit the comparison between the model and mock power spectra, and instead only show the recovered power spectra in Fig.\,\ref{fig:chain_cross_corr_sub_no_noise} (top). \citet{Kern+19}'s SVD method performs well and is able to essentially remove the cross-coupling systematics completely. Our sampling method performs similarly, although our model results in a fair amount of residual systematics, while \citet{Kern+19}'s is more prone to oversubtraction. To produce our result, a different model initialisation was required. Previous results all performed best with the ADVI initialiser (discussed in Section \ref{section:Sampler}). When modelling the cross-coupling in this low-noise cross-correlation, the systematics could only be mitigated by around two to four orders of magnitude, leaving a significant level of residual systematics. Opting instead for the {\tt jitter+adapt\_diag} initialiser improved this significantly. Here, a mass matrix for the HMC is formed, and a uniform jitter is applied to the starting sampler's points. This is an overall simpler method of initialisation in comparison to ADVI, and it is not clear to us why performance is worse for the preferred initialiser.

The mitigation of the strongest cross-coupling peaks appears somewhat easier for our model, given their strength and relative lack of confusion. The weaker, more confused peaks show poorer mitigation, resulting in these residual systematics. Nevertheless, the model performs well in this regime.

In terms of signal loss, both our and \citet{Kern+19}'s SVD approach perform similarly in terms of magnitude, although as mentioned, our method has a predisposition to incomplete mitigation, while \citet{Kern+19}'s tends to oversubtract more frequently. 

\subsection{Alternate/Incorrect Models}
We briefly discuss the results when the model for the mock visibilities is incorrect. Our fiducial mock data consists of 20 subreflections and 10 coupling peaks. We evaluate the results when: 
\begin{enumerate}
 \item Our fitting model contains too few or too many of these systematic features. Specifically, we test the cases of there being either 17 or 23 subreflections, and the cases of there being either 8 or 12 coupling peaks in our model. 

 \item There is the correct number of systematics features, but the parameters for them have been randomised to a degree, i.e. little effort has been put into making accurate initial estimates.
\end{enumerate}
Overall, the results are fairly consistent regardless of whether too many, too few, or randomised systematics features are assumed. All cases result in an excess of power following systematics mitigation, rather than an oversubtraction. Estimates of the systematics which are `near enough' to the true values result in slight suppression, while very poor estimates result in an addition of power. This is a result of the exponential terms in the systematics equations, Eqs.\:\ref{eq:CableReflectionGain} and \ref{eq:CC}, which themselves contain phases. For poor delay estimates, in particular, subtraction of model systematics does not lead to suppression of power, but rather addition.

\begin{figure}
    \centering
    \includegraphics[width=0.9\columnwidth]{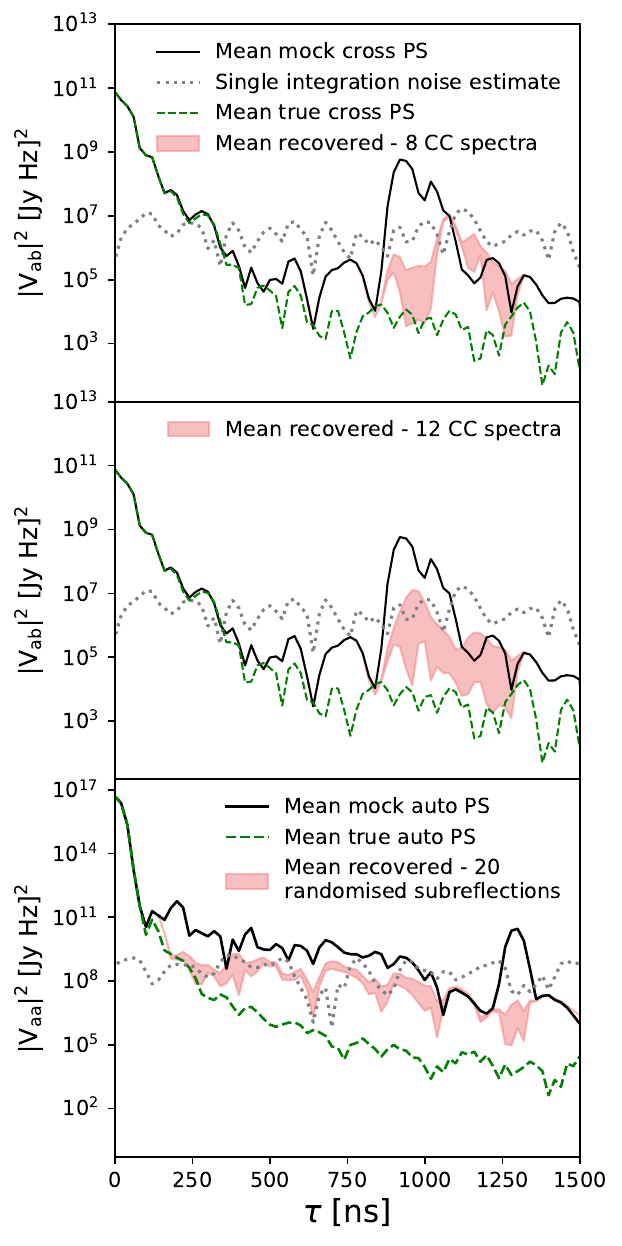}
    \caption{Examples of the mean recovered power spectra when the assumed systematics model differs from that present in the mock data. The fiducial coupling model contains ten peaks. The top plot shows the incoherently averaged recovered cross-correlation power spectra when only eight peaks are assumed, and the middle plot when twelve are assumed. The bottom plot shows the averaged recovered autocorrelations when the correct number of subreflections are assumed, but the initial delay estimates are randomly chosen. Similar results are observed across the different scenarios not shown here, i.e. too many/too few/randomised systematics in either the autocorrelations or cross-correlations.}
    \label{fig:8_vs_12_vs_rand}
\end{figure}

Both reflection modelling in the autocorrelations and coupling modelling in the cross-correlations show the same general behaviour for scenario (i). When there are too few systematics features, e.g. Fig. \ref{fig:8_vs_12_vs_rand} (top), the recovered power spectra exhibit excess power at the delays of the unmodelled systematics ($\sim1100-1200$\:ns). When there are too many assumed systematics, this produces similar excesses in power. Fig. \ref{fig:8_vs_12_vs_rand} (middle) shows an excess of power in the recovered spectra at delays where two too many coupling peaks are introduced ($\sim900-1000$\:ns).

Assuming the correct number of systematics in the correct delay range, but randomly placing the initial systematics peaks (with wider priors to account for this) produces somewhat more reasonable results, but still results in insufficient recovery. For most delays, and in the autocorrelations, recovery can be made down to the noise level for single time integrations when the subreflection peaks are randomly placed between $\sim200-1000$\:ns. However, attempting to incoherently average the recovered power spectra, as in Fig. \ref{fig:8_vs_12_vs_rand} (bottom), does not produce any meaningful level of noise reduction, suggesting there are significant amounts of residual subreflection systematics.

With randomised initial systematics parameters in the negligible noise scenario (not shown here), we see little to no mitigation of the systematics. This is mainly due to the low level of the noise, which results in the sampler less efficiently exploring the posterior space. Nevertheless, as before, we do not see any oversubtraction or signal loss, but this does suggest that a fair amount of effort should be put into setting the priors. This is not unlike the requirements of \citet{Kern+19}'s reflection mitigation method, which requires estimates of the reflection peak delay. However, \citet{Kern+19}'s coupling mitigation does have an advantage over ours, as only the delay range of the coupling peaks is required, whereas we again need initial delay, amplitude, and phase estimates.

Generally, poor priors for the systematics in our model are unlikely to result in oversubtraction, and are instead prone to induce excess power in the recovered spectra. It is not immediately obvious how to distinguish from the recovered spectra if there are too many assumed systematics features in the model versus too few, but it was noticed that when too few were assumed, the excess power demonstrated more of a typical peak shape. Assuming too many features resulted in wider confidence intervals in the region of the additional features.

\section{Discussion}\label{section:Discussion}
In this work, we present statistical modelling results when applied to the reflection and cross-coupling systematics in simulations of early HERA Phase I visibilities. These systematics serve to complicate the setting of upper limits on the EoR 21-cm signal by spreading copies of the foregrounds into regions ideally only occupied by the signal. We mimic this by adding systematics with randomised parameters to visibilities containing foregrounds and the 21-cm signal, where the initial, uncorrupted visibilities are sourced from the HERA Validation data \citep{Aguirre+22}, and the systematics are added with \textsc{hera\_sim}. A Hamiltonian Monte Carlo sampler is used to constrain these systematics, subject to noise, and attempts are made to remove them and recover the true foreground + EoR signal. We consider both an autocorrelation case, as well as a short, 14.7\,m east-west cross-correlation.

We test our model with and without an included signal model, and find that results are similar, though the signal-free model is preferred for its gains in sampling speed. At best, including a signal model provides slightly narrower uncertainty estimates in some cases, but this can lead to prohibitive runtimes, especially when noise is low or negligible.

The primary limitation of our method is the confusion between systematics, where nearly overlapping features makes it difficult for the model to accurately constrain the parameters, leading to residual systematics. This is most evident in cases where the noise is negligible, and more so for subreflection modelling in the autocorrelations, where multiple systematics peaks can occupy a small delay range. This is also a limitation of \citet{Kern+19}'s method, the currently employed calibrator in the HERA pipeline.

For autocorrelations, the twenty subreflections spread between $\sim200-1000$\:ns, and the coupling peaks at $\sim900-1300$\:ns, likely makes it difficult for the sampler to disentangle the power contributions from the multiple overlapping components, resulting in significant residual systematics in visibilities with low noise (Fig. \ref{fig:chain_auto_sub_no_noise}). To reduce this effect would likely require very specific priors on the systematics, something which is not always feasible given slight variations in array geometry, noise, RFI, and other complications in the observations. 

For cross-correlations, only ten cross-coupling spectra are modelled. Furthermore, these are high power relative to the foreground peaks, and are relatively unconfused, resulting in only slight residual systematics in low noise visibilities (Fig. \ref{fig:chain_cross_corr_sub_no_noise}). However, we only consider these unrealistic, essentially noise-free scenarios in order to evaluate signal loss. 

With higher noise levels, the model is less confusion-limited, as a significant portion of the systematics are either equivalent in power to, or below, the noise level. This leaves many of these systematics poorly constrained, which in this case is a non-issue, as this still allows for mitigation down to this high noise level (although further averaging/integration could cause them to emerge again). For low noise, these systematics require good constraints in order to mitigate them to any significant degree. For stronger, more isolated systematics like high-amplitude cable reflections, this is relatively easy to achieve, but for the more numerous, closely-spaced systematics (e.g. subreflections and coupling), finding good constraints is more difficult.

The difficulty in modelling the systematics in the low noise scenario is further exacerbated by the reduction in computational speed as a result of this noise level, which hinders quick tests of priors and sampler hyperparameter settings, as well as the model's ability to efficiently explore the posterior space. 

In autocorrelation power spectra, cable reflections, cable subreflections, and cross-coupling are of concern. These are analysed in Section \ref{subsection:AutocorrelationSystematics}. When considering a single 10.73\:second time integration, we are able to constrain a number of high-amplitude reflection systematics (Fig. \ref{fig:chain_auto}), and can mitigate down to the (relatively high) noise level. We apply this model to thirty time integrations separated by 21.46 seconds, and the recovered power spectra are incoherently averaged across time (Fig. \ref{fig:Auto_Recovered_mean}). This allows for a reduction in the noise level following systematics mitigation. These reflection systematics are modelled in the autocorrelations and applied to their lower-amplitude counterparts in the corresponding cross-correlations. For the same time integration, however, reflections in cross-correlations are overwhelmed by the noise by around four orders of magnitude. As such, there is not much to be gained from calibrating out reflections in the cross-coupling in this high noise case. 

This suggests that autocorrelations do not require complete noise removal in order to constrain the systematics, as even though the systematics mitigation is not perfect in this regime, it is ultimately the cross-correlations which are of scientific interest. It is here where noise would need to be minimised as much as possible. In other words, most of the analysis effort can be focused on the averaging and cleaning of cross-correlations, which requires a significant amount of data. Reflection systematics can be modelled in the autocorrelations, which themselves could perhaps be time averaged to a lower cadence, or undergo fewer or less intensive calibration, flagging, or cleaning steps, for example. The results from the autocorrelation systematics calibration can then be applied to the cross-correlations, with the only drawback being broader statistical uncertainty estimates.

When the model is run on visibilities with very low noise levels, however, the systematics in the autocorrelations can be mitigated such that the recovered power spectrum is, at worst, around three orders of magnitude above the true power spectrum (Fig. \ref{fig:chain_auto_sub_no_noise}). While this is extreme, using these results to subtract the reflections from the corresponding low noise cross-correlation results in essentially complete recovery of the true power spectrum in Fig.\,\ref{fig:chain_cross_sub_no_noise} (top). In this low noise case, there is minimal signal loss or residual systematics (Fig.\,\ref{fig:chain_cross_sub_no_noise}, bottom). Furthermore, as a result of the low noise, the constraints placed on the systematics parameters become very strong, resulting in narrow statistical uncertainty estimates.

In both the high noise and low noise cases, the non-linear optimisation method of \citet{Kern+19} performs similarly to our sampler when mitigating reflections. In the former case, systematics are mitigated to the noise level, and the noise can be further reduced by incoherently time averaging. In the latter case there are some residual systematics in the autocorrelation power spectrum, but there is very good recovery of the cross-correlation with minimal signal loss. It is in the cases where noise is more realistic that the value of a Bayesian approach is more evident, as it provides some measure of statistical uncertainty in the results. For lower noise levels, these estimates become narrower than is likely useful.

Only cross-coupling is modelled in cross-correlations (Section \ref{subsection:CrosscouplingSystematics}), as this systematic is significantly stronger here than in the autocorrelations, and it is sufficient to model the reflections in the single-antenna regime. Unfortunately, noise is also much stronger in the cross-correlations relative to the foreground peak.

For a single time integration, the strongest cross-coupling peaks are about as strong as the noise, eliminating any possibility of meaningful constraints being placed the systematics. Instead, some amount of noise reduction was required. Assuming a simple time integration increase from $\Delta t = 10.73\,\rm{s}$ to $1073\,\rm{s}$ reduces the noise enough to allow constraints to be placed on the highest amplitude cross-coupling peaks in Fig. \ref{fig:chain_cross_corr_tx100}, leading to their mitigation down to the noise level. This noise reduction is not representative of the methods used in HERA, but we take this approach to evaluate the model in the relatively high noise regime.

Once again mitigating the coupling systematics in thirty integrations separated by 21.46 seconds and incoherently averaging results in a further reduction of the noise level (Fig. \ref{fig:Cross_Recovered_mean}). As with other results, our model performs comparably to the SVD method of \citet{Kern+19}, within our uncertainty estimate. This required a different implementation of the sampler. All other results were found by sampling for a single visibility at a time. This result required a joint modelling approach, wherein the likelihood function had thirty terms, one for each time integration. All of these terms shared the same coupling systematics parameters as they are time-stable. By simultaneously modelling on additional data, the estimates on the parameters improved enough to allow for not only mitigation to the noise level, but further reduction of the noise after averaging, suggesting that the level of residual systematics in the recovered data was not significant.

When we evaluate a very low noise case in order to evaluate the model in an ideal scenario, i.e. where the noise power spectrum is around $10\%$ that of the EoR power spectrum, we are able to mitigate the cross-coupling spectrum down to, at worst, an order of magnitude above the true power spectrum, with there being recovery at a fair number of delays (Fig.\,\ref{fig:chain_cross_corr_sub_no_noise}, top). 

In this low noise modelling of the coupling systematic, \citet{Kern+19}'s SVD method performs similarly to our result. Their approach to deriving a cross-coupling model from the mock/observational data results in essentially complete mitigation. Both methods show similar levels of signal loss and residual systematics in terms of absolute dex (Fig.\,\ref{fig:chain_cross_corr_sub_no_noise}, bottom), although our model is more prone to undersubtraction, while \citet{Kern+19}'s to oversubtraction.

Hamiltonian Monte Carlo samplers are effective at modelling and mitigating reflection and cross-coupling systematics, while also providing a statistical uncertainty estimate of the recovered foreground and EoR signals. By modelling corrupted visibilities containing realistic levels of noise, the systematics can be subtracted to this noise level, and incoherently averaging the recovered spectra can further supress this noise. In essentially noise-free cases, true cross-correlation power spectra can be recovered with minimal signal loss and residual systematics, but the statistical uncertainty estimates are often narrower than is likely useful. In most cases, however, sampling can provide similar performance to the current systematic mitigation techniques, while remaining resilient to signal loss and providing statistical uncertainty measures. This can be important in both placing upper limits on, and possibly directly detecting, the early-Universe 21-cm power spectrum should the noise be reduced sufficiently. Performance is very dependent on a number of factors, however, such as priors and initialisation, particularly in low noise cases.

For HERA, the signal-chain systematics subtraction takes place after the majority of the analyses, namely redundant calibration, absolute calibration, RFI flagging, in-painting, etc. Only coherent time-averaging takes place after systematics removal, and thereafter power spectra are formed \citep{Aguirre+22}. As such, implementing a method similar to what is presented here would be fairly straightforward. Following systematics removal, instead of a single visibility for each time and each antenna, there would be however many sample visibilities. One option is to pass all of these visibilities through the pipeline and thousands of power spectra could be formed for each time and each antenna, from which uncertainties could be estimated. A less computationally expensive alternative is to form the uncertainties on the recovered visibilities, as was done in this work, and propagate those through the pipeline. For example, one could replace the observed visibilities with the upper and lower bounds of the percentile regions, effectively only doubling the amount of data, rather than increasing it by $\orderof(10^5)$ or more.

While the signal chain systematics presented here will mostly no longer be an issue in HERA Phase II, this work was meant to demonstrate that similar systematics which lend themselves to forward modelling can likely be mitigated in a similar manner. In Phase II, the cables connecting the FEM to the PAM are replaced by ones with a length of 500\:m, which pushes the reflections towards higher delays. The systems associated with the leaking connection point responsible for the cross-coupling are no longer in use, and so this systematic is no longer of concern. However, the introduction of Vivaldi feeds, and the lack of cages surrounding them, will likely result in over-the-air, antenna-to-antenna mutual coupling. Should this systematic be relatively time stable, and be capable of forward modelling, then the techniques presented in this work may be useful for their mitigation.

This work has a number of opportunities for further development. For the realistic noise cases, reflection modelling in the autocorrelations takes approximately 60 minutes for each individual time snapshot, given the large number of parameters and the use of the \textsc{ADVI} initialiser. This time includes model initialisation for two chains, a total of 8,000 tuning steps, and 10,000 sampling steps. Coupling modelling in a single cross-correlation, and for a single integration, is only of the order 10 minutes due to the reduced number of parameters and the use of a less computationally expensive initialiser. However, multiple integrations are included in the likelihood function for better results. Jointly modelling thirty integrations with two chains for a total of 10,000 tuning steps and 10,000 sampling steps takes approximately 3.5 hours.

For visibilities with negligible noise, run times extend into multiple hours for an individual time snapshot with similar settings. While levels of noise this low are not currently obtainable, for future data this would be an unreasonably high runtime. This is mainly due to the assumed noise level itself, where less noise typically results in slower sampling. 

While we fully intended to model each of these components, inferring them directly from mock/observational data, as is done for cross-coupling in \citet{Kern+19}, and thereafter adjusting them with a Bayesian approach could be an alternative. Good knowledge of the foreground components was also assumed beforehand, as the focus was placed on assessing the feasibility of modelling the systematics in a Bayesian framework. The effects of poor or incorrect models for the foregrounds and signal, and how this affects the mitigation of the systematics is another avenue which could be explored. Lastly, this work assumed a particular model for the sources of these systematics. It is possible that alternative models could result in an improvement in the systematics mitigation, and testing different models against one another is something Bayesian methods are ideally suited to.


\section*{Acknowledgements}

Analysis, writing, and editing was carried out by GGM, PB, and MGS. Further analysis of results was carried out by PK and JSD. The remaining authors, listed in alphabetical order, are members of the HERA Builder's List, which recognises their direct contributions to the construction, commissioning, and operation of the telescope and analysis pipeline development that are a necessary basis of the HERA-specific data and specifications used within this paper. 

We thank the anonymous reviewer for their feedback, which has resulted in significant improvement to the paper.

GGM and MGS acknowledge support from the South African National Research Foundation (Grant No. 84156).
This result is part of a project that has received funding from the European Research Council (ERC) under the European Union's Horizon 2020 research and innovation programme (Grant agreement No. 948764; P.~Bull). P.~Bull acknowledges support from STFC Grants ST/T000341/1 and ST/X002624/1.
We acknowledge the use of the ilifu cloud computing facility – www.ilifu.ac.za, a partnership between the University of Cape Town, the University of the Western Cape, Stellenbosch University, Sol Plaatje University and the Cape Peninsula University of Technology. The ilifu facility is supported by contributions from the Inter-University Institute for Data Intensive Astronomy (IDIA – a partnership between the University of Cape Town, the University of Pretoria and the University of the Western Cape), the Computational Biology division at UCT and the Data Intensive Research Initiative of South Africa (DIRISA).
We acknowledge the use of the following software: \textsc{numpy} \citep{Harris_2020}, \textsc{matplotlib} \citep{Hunter_2007}, \textsc{scipy} \citep{Virtanen_2020}, \textsc{pyuvdata} \citep{Hazelton_2017}, and \textsc{uvtools} (\href{https://github.com/HERA-Team/uvtools}{https://github.com/HERA-Team/uvtools}).

\section*{Data Availability}

This paper makes use of simulated data from \citet{Aguirre+22}. The analysis scripts used in this work are available from \href{https://github.com/GeoffMurphy/HMCSystematicsSampler}{https://github.com/GeoffMurphy/HMCSystematicsSampler}.

\balance


\bibliographystyle{mnras}
\bibliography{main}

\bsp	
\label{lastpage}
\end{document}